\documentclass[12pt]{iopart}
\usepackage{graphicx}
\usepackage{slashed}
\usepackage{multirow,array}
\usepackage{tikz}
\usetikzlibrary{decorations.markings}

\makeatletter
\newbox\slashbox \setbox\slashbox=\hbox{$/$}
\newbox\Slashbox \setbox\Slashbox=\hbox{\large$/$}
\def\pFMslash#1{\setbox\@tempboxa=\hbox{$#1$}
	\@tempdima=0.5\wd\slashbox \advance\@tempdima 0.5\wd\@tempboxa
	\copy\slashbox \kern-\@tempdima \box\@tempboxa}
\def\pFMSlash#1{\setbox\@tempboxa=\hbox{$#1$}
	\@tempdima=0.5\wd\Slashbox \advance\@tempdima 0.5\wd\@tempboxa
	\copy\Slashbox \kern-\@tempdima \box\@tempboxa}

\def\miss#1{\ifmmode{/\mkern-11mu #1}\else{${/\mkern-11mu #1}$}\fi}
\makeatother

\begin{document}

\title[CPT-Odd effects on the electromagnetic properties of charged leptons in...]{CPT-Odd effects on the electromagnetic properties of charged leptons in the Standard Model Extension}

\author{ J. S. Hurtado-Silva${}^{1}\label{Aut1}$, J. J. Toscano${}^{1}\label{Aut2}$,\\ and O. V\' azquez-Hern\' andez${}^{2}\label{Aut3}$}

\address{${}^{1}\ref{Aut1}$Facultad de Ciencias F\'isico Matem\'aticas, Benem\'erita Universidad Aut\'onoma de Puebla, Apartado Postal 1152 Puebla, Puebla, M\'exico\\
${}^{2}\ref{Aut3}$Universidad Michoacana de San Nicolás de Hidalgo. Av. Francisco J. Múgica s/n, C. P. 58060, Morelia, Michoacán, México}
\ead{juan.vazquezhurtado1@alumno.buap.mx, jtoscano@fcfm.buap.mx and oscar.vazquez@umich.mx}
\vspace{10pt}

\begin{abstract}
The impact of the CPT-Odd electroweak gauge sector of the Standard Model Extension on the electromagnetic properties of charged leptons is studied. This gauge sector is characterized by the $(k_1)_\mu$ and $(k_2)_\mu$ Lorentz violation (LV) coefficients, which have positive mass dimension because they are associated with a $U_Y(1)$-invariant and with an $SU_L(2)$-invariant dimension-three operators, respectively. They belong to the category of relevant interactions, which can have strong effects on low-energy observables. We present a comprehensive study on the impact of this sector on the magnetic dipole moment (MDM) and the electric dipole moment (EDM) of charged leptons, up to second order in these LV coefficients, both at the tree and one-loop levels. We find that the $O(k_i)$ contributions at the tree and one-loop levels depend on energy, while $O(k^2_i)$ ones at the tree-level do not. As for $O(k^2_i)$ one-loop effects, there are both energy-dependent and energy-independent contributions, but we have focused only on those of the latter type. We find that the EDM only is generated at $O(k_i)$ up to one-loop level, whereas the MDM receives contributions from both $O(k_i)$ and $O(k^2_i)$ at both tree and one-loop levels. The contributions of $O(k_i)$ to the MDM are found to be suppressed relative to the corresponding contributions to the EDM by approximately three orders of magnitude. Using a recent experimental limit on the electron EDM the $|(k_2)_0-|\mathbf{k_2}|\cos\theta_\gamma|<0.86\, m_e$ bound was obtained. As far as the contributions of $O(k^2_i)$ are concerned, we find that the tree-level contributions are suppressed with respect to the one-loop ones by at least a factor of $\left(m^2_l/m^2_Z\right)$. We find that the contribution to the electron MDM is by far the dominant one, as it can be up to four and seven orders of magnitude greater than those of the muon and tau, respectively. The Lorentz coefficient $(k_{AF})_\mu$ of the Carroll-Field-Jackiw's QED is given by a linear combination of the $(k_1)_\mu$ and $(k_2)_\mu$ vectors. Assuming that $|k^2_1|, |k^2_2|\gg |k^2_{AF}|$ and taking  $(k_{AF})_\mu=0$, which implies that $(k_1)_\mu$ and $(k_2)_\mu$ are collinear, we obtain an upper bound of $\left|\frac{ k^2_2}{m^2_e} \right|<4.36\times 10^{-10}$. The fact that $k^2_2$ is an observer Lorentz invariant allows us to introduce a new-physics scale through $\sqrt{k^2_2}=\Lambda_{CPT}$, for which we obtain the upper limit $\Lambda_{CPT}< 2.08 \times 10^{-5}\, m_e$. The physical implications derived from the fact that the LV coefficients have positive mass units are discussed. \\

\end{abstract}

%
%
%
%
%

\section{Introduction}\label{Int}
Over the past two decades, the idea that Lorentz and CPT transformations are not exact symmetries of nature in the very high energy domain has been the subject of significant interest in the literature. Diverse well-motivated theories already contain clues that suggest this possibility~\cite{QTvsGR,STLV,NCGLV,LQGLV,OCLV}. Therefore, it is important to look for signs of Lorentz violation (LV) effects at low energies. At low energies, the effects of Lorentz and CPT violation can be described in a model-independent way by the standard model extension (SME), which is an effective field theory that contains general relativity and the standard model (SM)~\cite{SMEGR,SME}. In its minimal version (mSME)~\cite{SME}, the model contains only renormalizable interactions, in the sense of mass units, but nonrenormalizable interactions are expected to play a dominant role at higher energies~\cite{NRSME,NR1,NR2,NR3,NR4}. The Lagrangian of the mSME is of the form:
\begin{equation}
	\label{SMEL}
	{\cal L}_{SME}={\cal L}_{SM}+\Delta {\cal L}\, , \end{equation}
where ${\cal L}_{SM}$ is the SM Lagrangian, while $\Delta {\cal L}$ is the part that introduces LV. The $\Delta {\cal L}$ Lagrangian contains all possible interactions of dimension less than or equal to four. To introduce LV into the theory, it is necessary to distinguish between two classes of Lorentz transformations, namely observer Lorentz transformation (OLT) (which corresponds to a passive transformation) and particle Lorentz transformation (PLT) (which corresponds to an active transformation). The $\Delta {\cal L}$ Lagrangian is constructed under the criterion that it is invariant under OLT, but not under PLT. According to this, the $\Delta {\cal L}$ Lagrangian is made of a sum of terms that are the product of two pieces, that is,
\begin{equation}
	\label{TO}
	T^{\mu_1\cdots \mu_n}{\cal O}_{\mu_1 \cdots \mu_n}\, ,
\end{equation}
where $T^{\mu_1\cdots \mu_n}$ represent constant Lorentz coefficients, while ${\cal O}_{\mu_1 \cdots \mu_n}$ is a function on the SM fields, which is gauge invariant and has canonical dimension less than or equal to four. Under OLT, both the $T^{\mu_1\cdots \mu_n}$ and ${\cal O}_{\mu_1 \cdots \mu_n}$ tensors transform covariantly, so terms of the way (\ref{TO}), and therefore $\Delta {\cal L}$, are invariant under this  type of transformation. However, under PLT, only the degrees of freedom represented by ${\cal O}_{\mu_1 \cdots \mu_n}$ are transformed, but $T^{\mu_1\cdots \mu_n}$ is not transformed, so $\Delta {\cal L}$ is not invariant under this type of transformation.

The SME contains some terms that are odd under CPT transformations. In the mSME, CPT-Odd interactions arise in both the fermion and boson sectors and have the peculiarity that the corresponding Lorentz coefficients $T^{\mu_1\cdots \mu_n}$ correspond in this case to  four-vectors $T^\mu$, which have positive mass dimension. In general, an effective field theory is characterized by a Lagrangian which contains unknown coefficients that multiply interactions constructed out with the SM fields. These coefficients can have positive mass dimension, be dimensionless or have negative mass dimension, which appear multiplying interactions of dimension less than four, of dimension four, and of dimension greater than four, respectively. In the order aforementioned, these interactions are usually classified as relevant, marginal and irrelevant. Interactions relevant or irrelevant imply in themselves a new physics scale given by their coefficients. Effects of relevant interactions on some observable are proportional to positive powers of $\Omega/M$, with $\Omega$ and $M$ the new physics scale and the natural scale of the process in consideration, respectively. In contrast the contributions of irrelevant interactions are proportional to positive powers of $M/\Lambda$, whit $\Lambda$ the new physics scale. Accordingly, relevant interactions  can give rise to effects which become quite large at energies much smaller than the scale of these coefficients. In contrast irrelevant interactions are small at low energies, but they become large at higher energies. On the other hand, marginal interactions are, in principle, important at all energy scales. The mSME is renormalizable, so it does not have irrelevant interactions, although, as already commented, the model can be enlarged to include them~\cite{NRSME,NR1,NR2,NR3,NR4}. In the mSME, all effective interactions are classified into CPT-Odd and CPT-Even. It turns out that, in this minimal version  of the model, the CPT-Odd interactions are relevant because they have dimension three, so their corresponding Lorentz coefficients have positive mass dimension. On  the other hand, the CPT-Even sector of the model contains only marginal interactions, so their Lorentz coefficients are dimensionless. As we will see in this paper, the fact that the CPT-Odd Lorentz coefficients have positive mass dimension has very interesting consequences on low-energy observables.

 In this paper, we are interested in studying the implications of CPT-Odd effects on the electromagnetic properties of charged leptons. We focus on the electroweak gauge sector of the mSME that is odd under CPT transformations. Since the MDM and the EDM of charged leptons are standard observables (in the sense that they are invariant under both OLT and PLT), the CPT-Odd contribution arises up to second order in the Lorentz coefficients, that is, the CPT-Odd effect is proportional to $T^2=T_\mu T^\mu$. Although standard observables do not carry information on spatial directions or relative motion, they can provide useful information about the importance of LV effects when constrained from high precision experiments, such as, for example, magnetic and electric dipole moments of charged leptons and nucleons~\cite{OP1,OP2,OP3}. Contributions to the MDM and EDM of charged leptons can also be generated through first-order effects in the LV coefficients~\cite{RR1,RR2}. Although these types of effects may depend on energy~\cite{RR1}, their study is interesting because they could provide information about spatial orientation or relative motion. In this paper, we will present a comprehensive study of these effects up to the second order in the LV coefficients, both at the tree and one-loop levels. Our purpose in this work is twofold. On the one hand, we are interested in deriving bounds on CPT-Odd effects from precision measurements and, on the other hand, in studying some interesting aspects of CPT-Odd effects at the level of radiative corrections~\cite{TV}. As we will see below, the CPT-Odd effects in radiative corrections on standard observables differ substantially from what occurs in any conventional extension of the SM (theories respecting Lorentz and CPT symmetries).

 The rest of the paper has been organized as follows. In Sec.~\ref{EWS}, the structure of the electroweak gauge sector of the mSME that is odd under CPT transformations is discussed and the Feynman rules needed for the calculation are derived. In Sec.~\ref{Cal}, the CPT-Odd contributions to the MDM and EDM of a charged lepton at both tree and one loop levels are calculated. Contributions up to second order in the LV coefficients are considered. Sec.~\ref{Dis} is devoted to discussing our results. In Sec.~\ref{Con} the conclusions are presented. Details of calculations are presented in Appendices A, B, and C.

\section{The CPT-Odd electroweak gauge sector}
\label{EWS}
The CPT-Odd electroweak gauge sector of the mSME is given by:
\begin{eqnarray}
	\label{LW}
	{\cal L}^{CPT-Odd}_{W}&=&\frac{1}{2}(k_2)_\lambda \epsilon^{\lambda \rho \mu \nu}Tr\left(W_\rho W_{\mu \nu}+\frac{2}{3}igW_\rho W_\mu W_\nu \right)\, , \\
	\label{LB}
	{\cal L}^{CPT-Odd}_{B}&=&\frac{1}{2}(k_1)_\lambda  \epsilon^{\lambda \rho \mu \nu}B_\rho B_{\mu \nu}\, ,
\end{eqnarray}
where $W_\mu=\frac{\sigma^i}{2}W^i_\mu$ and $W_{\mu \nu}=\frac{\sigma^i}{2}W^i_{\mu \nu}$, with $W^i_{\mu \nu}=\partial_\mu W^i_\nu -\partial_\nu W^i_\mu+g\epsilon^{ijk}W^j_\mu W^k_\nu$, are the gauge fields and the curvatures associated with the $SU_L(2)$ gauge group, respectively. On the other hand, $B_\mu$ and $B_{\mu \nu}=\partial_\mu B_\nu-\partial_\nu B_\mu$ are the gauge field and curvature of the $U_Y(1)$ gauge group. The Lagrangians (\ref{LW}) and (\ref{LB}) are gauge-invariant up to a surface term. Notice that the $(k_2)_\mu$ and $(k_1)_\mu$ Lorentz coefficients have units of mass. As we will see below, this fact has important consequences in low-energy observables.\\

From the perspective of the electromagnetic $U_Q(1)$ gauge group, the (\ref{LW}) and (\ref{LB}) Lagrangians look like
\begin{eqnarray}
	\label{LWW}
	\fl {\cal L}^{CPT-Odd}_{W}=\frac{1}{4}(k_2)_\lambda \epsilon^{\lambda \rho \mu \nu}\left(W^+_\rho \hat{W}^-_{\mu \nu}+W^-_\rho \hat{W}^+_{\mu \nu}+
	W^3_\rho W^3_{\mu \nu}-\frac{2ig}{3}W^3_{[\rho} W^+_\nu W^+_{\mu]} \right)\, ,
\end{eqnarray}
\begin{equation}
	\label{LBB}
	{\cal L}_B=\frac{1}{2}(k_1)_\lambda \epsilon^{\lambda \rho \mu \nu}\left[c^2_W A_\rho F_{\mu \nu}+s^2_W Z_\rho Z_{\mu \nu}-s_Wc_W\left(A_\rho Z_{\mu \nu}+Z_\rho F_{\mu \nu} \right) \right]\, ,
\end{equation}
where the well-known map relating the gauge fields $(W^1_\mu, W^2_\mu, W^3_\mu, B_\mu)$ with the mass eigenstates fields $(W^-_\mu, W^+_\mu, Z_\mu, A_\mu)$ has been used. In the above expressions,
\begin{eqnarray}
	\hat{W}^+_{\mu \nu}&=&W^+_{\mu \nu} +ig(W^+_\mu W^3_\nu-W^+_\nu W^3_\mu)\, , \\
	\hat{W}^-_{\mu \nu}&=&\left(\hat{W}^+_{\mu \nu}\right)^\dag\, , \\
	W^3_{\mu \nu}&=&s_WF_{\mu \nu}+c_W Z_{\mu \nu}+ig\left(W^-_\mu W^+_\nu -W^+_\mu W^-_\nu \right)\, ,
\end{eqnarray}
where $W^+_{\mu \nu}=\partial_\mu W^+_\nu-\partial_\nu W^+_\mu$, $W^3_\mu=c_WZ_\mu +s_WA_\mu$, $Z_{\mu \nu}=\partial_\mu Z_\nu-\partial_\nu Z_\mu$, and $F_{\mu \nu}=\partial_\mu A_\nu-\partial_\nu A_\mu$. In Eq.(\ref{LWW}), the symbol $[\rho \nu \mu]$ represents the sum of cyclic permutations in the indices $\rho$, $\nu$ and $\mu$. In addition, $s_W(c_W)$ stands for sine(cosine) of the weak angle $\theta_W$.\\

\begin{figure}[h]
	\includegraphics[trim= -20mm 220mm 0mm 20mm, scale=0.70,clip, angle=0]{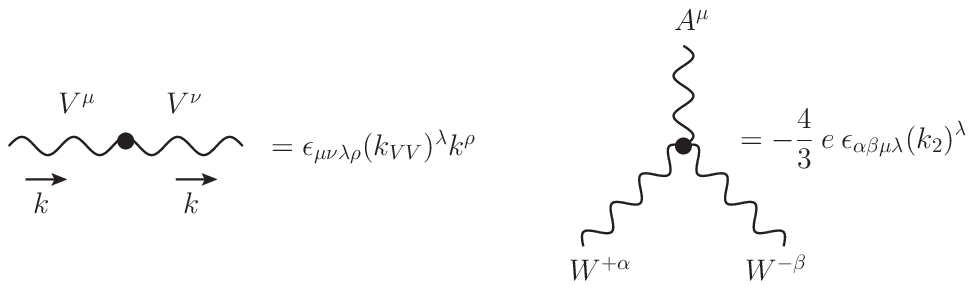}
	\caption{ Feynman rules arising from the CPT-Odd sector, which are needed for the calculation.}	\label{FR}
\end{figure}

The tree- and one-loop level contributions of the (\ref{LWW}) and (\ref{LBB}) Lagrangians to the electromagnetic properties of charged leptons are given by the $AA$, $AZ$, $ZZ$, and $W^-W^+$  insertions and for the $WW\gamma$ vertex. The corresponding couplings are given by the Lagrangians:
\begin{eqnarray}
	\label{CFJ}
	{\cal L}^{CPT-Odd}_{AF}&=&\frac{1}{4}(k_{AF})_\lambda \epsilon^{\lambda\rho \mu \nu}A_\rho F_{\mu \nu} \, , \\
	{\cal L}^{CPT-Odd}_{ZZ}&=&\frac{1}{4}(k_{ZZ})_\lambda \epsilon^{\lambda\rho \mu \nu}Z_\rho Z_{\mu \nu} \, , \\
	{\cal L}^{CPT-Odd}_{AZ}&=&\frac{1}{4}(k_{AZ})_\lambda \epsilon^{\lambda\rho \mu \nu}\left(A_\rho Z_{\mu \nu}+Z_\rho F_{\mu \nu}\right) \, , \\
	{\cal L}^{CPT-Odd}_{WW}&=&\frac{1}{4}(k_{2})_\lambda \epsilon^{\lambda\rho \mu \nu}\left(W^+_\rho W^-_{\mu \nu}+W^-_\rho W^+_{\mu \nu}  \right)\, , \\
	{\cal L}^{CPT-Odd}_{WW\gamma}&=&\frac{4ie}{9}(k_2)_\lambda \epsilon^{\lambda\rho \mu \nu}\left(A_\rho W^-_\mu W^+_\nu+A_\mu W^+_\rho W^-_\nu +
	A_\nu W^-_\rho W^+_\mu\right)\, ,
\end{eqnarray}
where
\begin{eqnarray}
	\label{CLAA}
	k_{AF}&=&2c^2_Wk_1+s^2_Wk_2\, , \\
	\label{CLZZ}
	k_{ZZ}&=&2s^2_Wk_1+c^2_Wk_2\, , \\
	\label{CLAZ}
	k_{AZ}&=&s_Wc_W(k_2-2k_1)\, .
\end{eqnarray}
The Feynman rules of the above two- and three-point vertex functions are shown in Fig~\ref{FR}.\\

The LV coefficient $k_{AF}$ given by Eq.(\ref{CLAA}), which is associated with the Carroll-Field-Jackiw's Lagrangian~\cite{CFJ}, has been the subject of considerable interest in the literature. In this pioneering work, data on cosmological birefringence were used to impose upper bounds of order of $10^{- 25} \, GeV$ and $10^{-42}\, GeV$, for a Lorentz coefficient of the form $(k_{AF})^\alpha=(k,\vec{0})$ and for a timelike $(k_{AF})^\alpha$, respectively. Recently, stronger upper bounds have been obtained from cosmic microwave background searches~\cite{Caloni}. In this work, an upper bound on the parameter $k^{(3)}_{(V)00}=-\sqrt{4\pi}(k_{AF})^0$ of $|k^{(3)}_{(V)00}|<1.54\times 10^{-44}\, GeV$ has been obtained. In the same work, a bound of $|\mathbf{k}_{AF}|<7.4\times 10^{-45}\, GeV$ on the space components of $(k_{AF})_\mu$ was obtained. Bounds on these parameters derived in previous literature are collected in table D15 of  reference~\cite{TK}.\\

Because of the very strong bounds on the Lorentz coefficient $(k _{AF})_\mu$, it makes sense to assume that this parameter is, for all practical purposes, equal to zero. This in turns implies from the relation given by Eq.(\ref{CLAA}) that the Lorentz coefficients $(k_1)_\mu$ and $(k_2)_\mu$ are collinear, that is,

\begin{equation}
	\label{cr}
	(k_1)_\mu=-\frac{1}{2}\frac{s^2_W}{c^2_W}(k_2)_\mu \, .
\end{equation}
This assumption will be used throughout this work.

\section{The calculation}
\label{Cal}As commented in the introduction, the presence of the four-vector $(k_i)_\mu$ (with $i=1,2$) leads to new electromagnetic form factors, which, even in zero momentum transfer, can depend on the energy. Such a dependence can arise through scalar products of the form $p\cdot k_i$, with $p$ an external momentum. As the order in $k_i$ considered increases, the diversity of electromagnetic gauge structures increases along with the possible energy dependence of the associated form factors~\cite{MNTT}. We illustrate this by focusing on magnetic form factors. We will consider up to second order effects on the vector $k_i$ at both the tree and one-loop levels. To this order in $k_i$, one can write a gauge invariant effective Lagrangian for the MDM with the following Lorentz structure,
\begin{eqnarray}
	\label{elmm}
	{\cal L}_{AMM}=&\frac{f_1}{4m_l}\,\left(\bar{l}\sigma_{\mu \nu}l\right) \,F^{\mu \nu}+\frac{f_2}{4m_l^3}\, k^\lambda_i \left(i\overline{D_\lambda l}\sigma_{\mu \nu}l -i\bar{l}\sigma_{\mu \nu}D_\lambda l\right)\, F^{\mu \nu}\nonumber\\	&+\frac{f_3}{4m^5_l}\, k^\lambda_i k^\rho_i\, \left(\overline{D_\lambda l}\sigma_{\mu \nu}D_\rho l\right)F^{\mu \nu}\, ,
\end{eqnarray}
where $D_\mu=\partial_\mu -ie A_\mu$ is the electromagnetic covariant derivative, $\sigma_{\mu \nu}=\frac{i}{2}[\gamma_\mu,\gamma_\nu]$, and $m_l$ the mass of the lepton $l$. In addition,  $f_i$ are real dimensionless functions on Lorentz scalars.\\

The MDM interaction induced by the Lagrangian (\ref{elmm}) is given by the following vertex function:
\begin{equation}
	\Gamma^{MDM}_\mu=\left(\hat{F}_1+\hat{F}_2+\hat{F}_3\right)\frac{i\sigma_{\mu \nu}q^\nu }{2m_l}\, ,
\end{equation}
where $q=p_1-p_2$ is the photon momentum and $p_1(p_2)$ the momentum of the ingoing(outgoing) lepton $l$. The form factor $\hat{F}_1$, which coincides with the $f_1$ function that appears in the Lagrangian Eq.(\ref{elmm}), can be decomposed as the sum of two parts as follows:
\begin{equation}
	\hat{F}_1=F^{SM}_1(q^2,m^2_l,m^2_V)+\frac{k^2_i}{m^2_l}\tilde{F}_1(q^2,m^2_l,m^2_V)\, ,
\end{equation}
where $k^2_i=(k_i)_\mu (k_i)^\mu$ and $m_V$ stands for the mass of the weak gauge bosons $m_Z$ or $m_W$. In this expression, $F^{SM}_1(0,m^2_l,m^2_V)=a^{SM}_l$ is the SM prediction for the MDM of the charged lepton $l$. As with the SM contribution, the form factor $\tilde{F}_1(0,m^2_l,m^2_V)$ is not energy dependent and, as we will show below, is a well-behaved function on the  $m_l$ and $m_V$ masses. This type of contribution is generated both at the tree and one-loop levels, but, as we will see below, the latter
dominates on the former.\\

Regarding the form factors $\hat{F}_2$ and $\hat{F}_3$, they are given by:

\begin{eqnarray}
	\hat{F}_2&=&\frac{k_i\cdot (p_1+p_2)}{m^2_l}\, f_2\, ,\\
	\hat{F}_3&=&\frac{(k_i\cdot p_1)(k_i\cdot p_2)}{m^4_l}\, f_3\, .
\end{eqnarray}
From these expressions, we can see that these form factors depend on energy. We will see later that $\hat{F}_2$ and $\hat{F}_3$ receive contributions at both the tree and one-loop levels. However, we will show that one-loop contributions dominate over tree contributions. A similar analysis can be given regarding the electric form factor. As we will see later, the EDM only receives contributions of $O(k_i)$ at the one-loop level.

\subsection{The tree level contribution}
\label{Tree} 
\begin{figure}[h]
	\includegraphics[trim= 0mm 245mm 0mm 20mm, scale=0.85,clip]{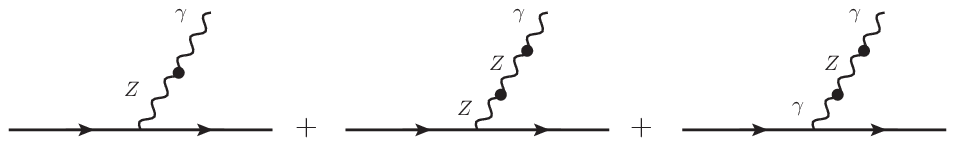}
	\caption{Tree-level diagrams contributing to the magnetic dipole moments of charged leptons at first- and second-order in the LV coefficients $(k_{1})_\mu$ and $(k_{2})_\mu$.  }\label{Tree-1-2}
\end{figure}

In the $(k _{AF})_\mu=0$ approximation, the tree-level contribution to the on-shell vertex $\bar{l}l\gamma$, with $l=e, \ \mu, \tau$, is given by the reducible diagrams shown in Fig.~\ref{Tree-1-2}. To first order in the Lorentz coefficients, the contribution is given by the first diagram of this figure. The amplitude in the presence of a static magnetic field can be written as follows:
\begin{equation}
	{\cal M}^{tree(1)}=-\frac{g}{2c_W}\frac{\epsilon_{\mu \lambda \rho \alpha}k^\lambda_{AZ}q^\rho}{q^2-m^2_Z}\bar{u}(p_2,s_2)\left(g^l_V+g^l_A\gamma_5\right) \gamma^\alpha \, u(p_1,s_1)A^\mu(q)\, .
\end{equation}
After performing some algebraic manipulations, we obtain
\begin{eqnarray}
	\fl {\cal M}^{tree(1)}=\frac{igg^l_A}{2c_W}\frac{1}{q^2-m^2_Z}\bar{u}(p_2,s_2)\left[(p_1+p_2)^\mu \pFMSlash{k}_{AZ}-(p_1+p_2)\cdot k_{AZ} \right]u(p_1,s_1)A^\mu(q)+\cdots\nonumber\\
\end{eqnarray}
where the ellipses indicate terms proportional to the $\gamma_5$ matrix. We have verified that these terms do not contribute to the EDM. Following Ref.\cite{RR2}, finally we get
\begin{equation}
	\label{tree1}
	a^{tree(1)}_l=\frac{(p_1+p_2)\cdot k_2}{4m^2_W}\, ,
\end{equation}
to zero transfer moment. In addition, Eqs.(\ref{CLAA}) and (\ref{CLAZ}) where used together with the assumption $(k_{AF})_\mu=0$. From this expression, it can be seen that the contribution to the MDM of the lepton $l$ depends on the energy and that it could only be relevant for energies of the order or higher than the Fermi scale.\\

Regarding the contribution at the tree level of order $k^2_i$, it is given by the second and third diagrams in Fig.~\ref{Tree-1-2}. The second diagram involves the exchange of  two virtual $Z$ bosons, while the third diagram involves a virtual photon and a virtual $Z$. The corresponding amplitude in the presence of a static magnetic field is given by:
\begin{eqnarray}
	{\cal M}^{tree(2)}&=&-\sum_{V=\gamma,Z}\frac{i v}{(q^2-m^2_V)(q^2-m^2_Z)}\bar{u}(p_2,s_2)\gamma_\alpha u(p_1,s_1)\nonumber \\
	&&\times \Big\{g^{\alpha \mu}\left[q^2(k_{AZ}\cdot k_{VZ})-(k_{AZ}\cdot q) (k_{VZ}\cdot q)\right]\nonumber \\
	&& \, \,  \, \, \, \, + q^\alpha \left[k^\mu_{VZ}(k_{AZ}\cdot q)-q^\mu (k_{AZ}\cdot k_{VZ})\right]\nonumber \\
	&& \, \,  \, \, \, \, + k^\alpha_{AZ}\left[q^\mu (k_{VZ}\cdot q)-q^2 k^\mu_{VZ}\right]\Big\} A_\mu(q)+\cdots \, ,
\end{eqnarray}
where $v=e$ for $V=\gamma$ and $v=\frac{g}{2c_W}g^l_V$ for $V=Z$. In addition, the ellipsis indicate terms proportional to the $\gamma_5$ matrix. We have verified that, as occurs in the contribution of $O(k_i)$, there is no contribution to the EDM. Noting that the $V=Z $ contribution is suppressed with respect to the $V=\gamma$ contribution by a factor $1/m^2_Z$, we will focus only on the latter contribution. In this approximation, we have
\begin{eqnarray}
	{\cal M}^{tree(2)}&=&\frac{ie}{q^2-m^2_Z}\bar{u}(p_2,s_2)\gamma_\alpha u(p_1,s_1)\Bigg\{  g^{\alpha \mu}k^2_{AZ}-k^\alpha_{AZ}k^\mu_{AZ}\nonumber \\
	&&+(k_{AZ})_\beta\left[k^\alpha_{AZ}\frac{q^\mu q^\beta}{q^2}-k^\mu_{AZ}\frac{q^\alpha q^\beta}{q^2}\right] \Bigg\}A_\mu(q)\, .
\end{eqnarray}
In the $q_\mu \to 0$ limit, the coefficient of $(k_{AZ})_\beta$ vanishes, so the contribution to the MDM is given by the term proportional to $k^2_{AZ}$. Using Eqs.(\ref{CLAA}) and (\ref{CLAZ}) together with the assumption $(k_{AF})_\mu=0$, it is found that such contribution is:
\begin{equation}
	\label{tree2}
	a^{tree(2)}_l= \frac{s_W}{c_W}\frac{k^2_2}{m^2_Z}\, .
\end{equation}
Note that this contribution does not depend on energy. However, as we will show later, the $Z$ exchange introduces a strong suppression with respect to the
one loop level contribution.

\subsection{The one-loop contribution}
\label{Loop}
\begin{figure}[h]
	\includegraphics[trim= 20mm 205mm 0mm 20mm, scale=1.0,clip]{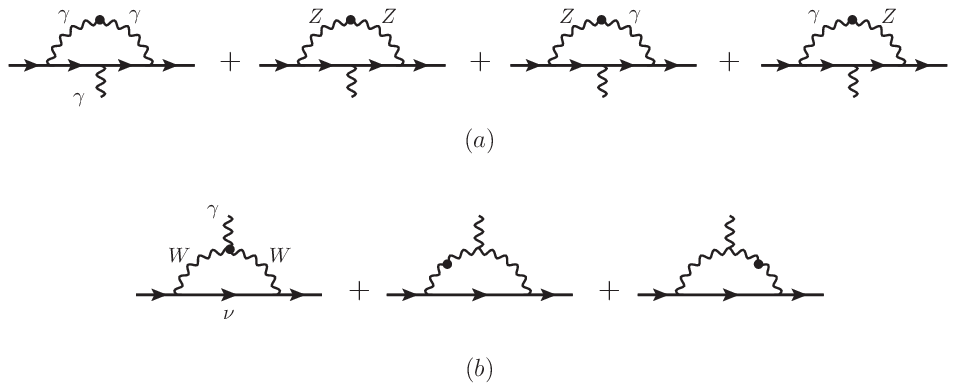}
	\caption{One-loop diagrams contributing to the magnetic dipole moment of charged leptons to first order in the Lorentz coefficient $(k_i)_\mu$.}\label{Loop1}
\end{figure}

\begin{figure}[h]
	\includegraphics[trim= 20mm 160mm 0mm 20mm, scale=1.0,clip]{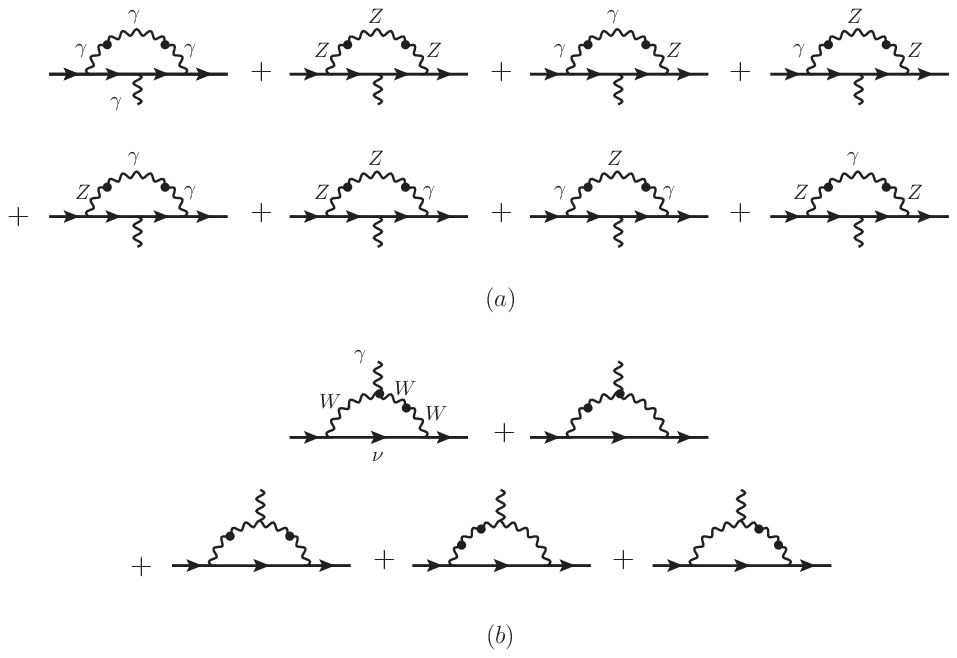}
	\caption{One-loop diagrams contributing to the magnetic dipole moment of charged leptons to second order in the Lorentz coefficient $(k_i)_\mu$.}\label{Loop2}
\end{figure}

We now proceed to discuss the CPT-Odd contribution to the MDM and EDM of the lepton $l$ at the one loop level. As in the tree level case, both first- and second-order contributions in the LV coefficients $(k_i)_\mu$ will be considered. These contributions are given through the Feynman diagrams shown in Figs.~\ref{Loop1} and \ref{Loop2}. As can be seen in these figures, the three electroweak gauge bosons contribute to the process that interests us, so it is important to verify if our results are independent of the procedure used to fix the gauge. All physical observable must be gauge-independent, i.e. not dependent on the gauge parameter $\xi$. To carry out this, we will proceed as in the tree-level case by working with the general propagators for these gauge bosons generated in the context of the $R_\xi$ gauges. The following fact simplifies the calculations considerably. In the $R_\xi$-gauge, the gauge-boson propagators are given by:
\begin{equation}
\label{GP}
	P^{\mu \nu}=\frac{-i}{k^2-m^2_V}\left(g^{\mu \nu}-(1-\xi)\frac{k^\mu k^\nu }{k^2-\xi m^2_V}\right)\, ,
\end{equation}
where $m_V$ stands for $m_Z$ or $m_W$. The photon propagator is obtained from this expression by putting $m_V=0$. Now, notice that all the diagrams in Figs.~\ref{Loop1}$(a)$ and \ref{Loop2}$(a)$ contain contractions between neutral gauge bosons, each of which involves the contraction of a two-point vertex function with its two adjacent propagators, that is,
\begin{equation}
	P^{\alpha \mu}\, \left[\epsilon_{\mu \nu \lambda \rho}(k_{VV})^\lambda k^\rho\right] \,P^{\nu \beta}=-\frac{\epsilon^{\alpha \beta \lambda \rho}(k_{VV})_\lambda k_\rho}{\left(k^2-m^2_{V}\right)^2}\, ,
\end{equation}
which shows us that only the transverse part of the propagators contributes to the amplitude. This means that in the unitary gauge, which corresponds to the $\xi \to \infty$ limit in the general propagator (\ref{GP}), the longitudinal modes of the $Z$ or $W$ gauge bosons do not contribute. In the case of the photon, this is equivalent to using the Feynman-'t Hooft gauge given by $\xi=1$. Furthermore, this result shows us that each diagram leads by itself to an amplitude that is gauge-independent and, as will become clear later, free of ultraviolet (UV) divergences. This means that it makes sense to analyze the relative importance of the contribution of each diagram to the MDM and EDM.\\

As far as the contribution of the $W$ gauge boson is concerned, it is given through diagrams shown in Figs.~\ref{Loop1}$(b)$ and \ref{Loop2}$(b)$. In the $R_\xi$-gauge, simplifications as those of diagrams of Figs.~\ref{Loop1}$(a)$ and \ref{Loop2}$(a)$ are also obtained. However, due to the presence of the SM $WW\gamma$ vertex in some of these  diagrams, two additional diagrams must be added when this vertex is replaced by the $WG_W\gamma$ vertex, with $G_W$ the pseudo-Goldstone boson associated with the $W$ gauge boson. The unphysical $WG_W\gamma$ vertex arises in the context of linear $R_\xi$ gauges, in which the gauge-fixing term does not affect the vertices of the theory but only its quadratic part, leading to the general propagator (\ref{GP}). However, one can introduce a nonlinear gauge-fixing procedure that fixes the gauge of the $W^\pm$ gauge boson in a covariant way under the electromagnetic gauge group $U_Q(1)$. This nonlinear $R_\xi$-gauge is defined through the following gauge-fixing functions~\cite{NLG}:

\begin{eqnarray}
f^+&=& D_\mu W^{+\mu}-\xi m_W G^+_W \, , \\
f^Z&=&\partial_\mu Z^\mu-\xi m_Z G_Z\, , \\
f^A&=&\partial_\mu A^\mu\, ,
\end{eqnarray}
where $D_\mu=\partial_\mu-ie A_\mu$ is the electromagnetic covariant derivative, $G^+_W (G_Z)$ is the pseudo-Goldstone boson associated with the $W^+(Z)$ gauge boson, and $f^-=( f^+)^\dag$. This nonlinear gauge-fixing procedure reduces to the linear one when the covariant derivative is replaced by the partial derivative. This type of gauge introduces nontrivial changes in the Higgs and Yang-Mills sectors of the electroweak theory. In addition to generating the general propagator (\ref{GP}), this type of gauge cancels the $WG_W\gamma$ vertex arising from the Higgs sector. Furthermore, this gauge substantially modifies the $WW\gamma$ vertex that arises from the Yang-Mills sector. In the non-linear gauge, the vertex function associated with the $A^\mu (k_1) W^{+\lambda}(k_2)W^{-\rho}(k_3)$ coupling is given by $ie\, \Gamma_{\lambda \rho \mu}(k_1,k_2,k_3)$, where:
\begin{equation}
\fl	\label{NLGV}
\Gamma_{\lambda \rho \mu}(k_1,k_2,k_3)=\left(k_2-k_3\right)_\mu g_{\lambda \rho}+\left(k_3-k_1+\frac{1}{\xi}k_2 \right)_\lambda g_{\rho \mu}
	+\left(k_1-k_2-\frac{1}{\xi}k_3 \right)_\rho g_{\lambda \mu}  \, ,
\end{equation}
where all momenta are taken incoming. As already mentioned, each diagram in Figs.~\ref{Loop1}$a$ and \ref{Loop2}$a$ independently contributes to the MDM and EDM, but this is not the case for the diagrams in Figs.~\ref{Loop1}$b$ and \ref{Loop2}$b$. This is more evident in the case of the diagrams in Fig. $4b$, since those diagrams characterized by the SM vertex $WW\gamma$ lead to individual contributions involving the matrix $\gamma_5$, appearing to contribute to the electric dipole moment. However, such contribution is spurious as it is canceled once all diagrams are added. To calculate these diagrams, we will use the Feynman-'t Hooft gauge ($\xi=1$). In this gauge, the vertex (\ref{NLGV}) takes a simple form:
\begin{equation}
	\Gamma_{\lambda \rho \mu}=2\left(-k_{2\mu}g_{\lambda \rho}+k_{1\lambda}g_{\rho \mu}-k_{1\rho}g_{\lambda \mu}\right)\, .
\end{equation}

At this point it is worth making some comments about the possible presence of UV and infrared (IR) divergences. From a direct analysis on the diagrams in Figs.~\ref{Loop1} and \ref{Loop2}, we can conclude that the amplitude associated with each of them is free of UV divergences, which is due to the unusual large amount of propagators involved. Regarding the presence of IR divergences, we have verified that this type of divergence only arises from the first diagram in Figs. $3(a)$ and $4(a)$, but they are absent in the $(k_{AF})_\mu=0$ approximation. It should be noted that, in this approximation, there are no contributions from the rest of diagrams containing photon-photon type insertions, although in these cases the associated amplitudes are free of IR divergences.\\

\subsubsection{$O(k)$ contribution}
\label{L1}
\ \\
To first order in the LV coefficients, the contribution to the MDM and EDM of the charged lepton $l$ is given by diagrams shown in Fig.~\Ref{Loop1}. Following Ref.~\cite{RR2}, the amplitude associated with the current generated by the $l$ lepton coupled to a static magnetic field can be written as follows:
\begin{eqnarray}
	\mathcal{M}=i e \, \bar{u}(p_{2},s_{2}) \, \Gamma_{\mu}^{SME}  \, u(p_{1},s_{1}) A^{\mu}(q)\, ,
\end{eqnarray}
where
\begin{eqnarray}
	\Gamma_{\mu}^{SME}= \Gamma_{\mu}^{SM}+ \Gamma_{\mu}^{\mathcal{O}(1)}\, ,
\end{eqnarray}
with  $\Gamma_{\mu}^{SM}$ the SM contribution and  $\Gamma_{\mu}^{\mathcal{O}(1)}$ the $O(k)$ CPT-Odd contribution. From now on, we focus only in the CPT-Odd contribution, which can be written as follows:

\begin{eqnarray}
	\Gamma_{\mu}^{\mathcal{O}(1)}=\Gamma_{\mu}^{(\gamma\gamma)}+\Gamma_{\mu}^{(ZZ)}+\Gamma_{\mu}^{(Z\gamma)}+\Gamma_{\mu}^{(\gamma Z)}+\Gamma_{\mu}^{(WW)}\, ,
\end{eqnarray}
where the first four terms correspond to the amplitudes of the first four diagrams in Fig.~\Ref{Loop1}$(a)$, whereas the last term corresponds to the sum of amplitudes associated with the diagrams of Fig.~\Ref{Loop1}$(b)$:
\begin{eqnarray}
	\label{WW}
	\Gamma_{\mu}^{(WW)}=\Gamma_{\mu}^{(WW)(1)}+\Gamma_{\mu}^{(WW)(2)}+\Gamma_{\mu}^{(WW)(3)}\, .
\end{eqnarray}
In this expression, $\Gamma_{\mu}^{(WW)(1)}$ is associated with the diagram containing the CPT-Odd $WW\gamma$ vertex, whereas the $\Gamma_{\mu}^{(WW)(2)}$ and $\Gamma_{\mu}^{(WW)(3)}$ amplitudes arise from the SM  $WW\gamma$ vertex. \\

The amplitudes associated with the diagrams in Fig.~\Ref{Loop1}$(a)$ are given by:
\begin{eqnarray}
\fl \Gamma_{\mu}^{(\gamma\gamma)}=-e^2 \int\frac{d^4k}{(2\pi)^4}\frac{\gamma_{\alpha}(\slashed{p_{2}}-\slashed{k}+m_{l})\gamma_{\mu}(\slashed{p_{1}}-\slashed{k}+m_{l})\gamma_{\beta} \, \epsilon^{\alpha\beta\sigma\tau}(k_{AF})_{\sigma}k_{\tau}}{\left[k^2-m_{\gamma}^2\right]^2\left[(k-p_{1})^2-m_{l}^2\right]\left[(k-p_{2})^2-m_{l}^2\right]}\,  ,\,\,\,\label{AA}
\end{eqnarray}
\begin{eqnarray}
\fl \Gamma_{\mu}^{(ZZ)}=-\frac{g^2}{4c_{W}^2}\int\frac{d^4k}{(2\pi)^4}\frac{\gamma_{\alpha}(g^{l}_{V}-g_{A}^{l}\gamma_{5})(\slashed{p_{2}}-\slashed{k}+m_{l})\gamma_{\mu}(\slashed{p_{1}}-\slashed{k}+m_{l})\gamma_{\beta}(g^{l}_{V}-g_{A}^{l}\gamma_{5}) }{\left[k^2-m_{Z}^2\right]^2\left[(k-p_{1})^2-m_{l}^2\right]\left[(k-p_{2})^2-m_{l}^2\right]}\nonumber\\
\times\epsilon^{\alpha\beta\sigma\tau}(k_{ZZ})_{\sigma}k_{\tau}\:  \label{ZZ}
\end{eqnarray}
\begin{eqnarray}
\fl \Gamma_{\mu}^{(Z\gamma)}=\frac{ge}{2c_{W}}\int\frac{d^4k}{(2\pi)^4}\frac{\gamma_{\alpha}(\slashed{p_{2}}-\slashed{k}+m_{l})\gamma_{\mu}(\slashed{p_{1}}-\slashed{k}+m_{l})\gamma_{\beta}(g^{l}_{V}-g_{A}^{l}\gamma_{5})\, \epsilon^{\alpha\beta\sigma\tau}(k_{AZ})_{\sigma}k_{\tau}}{\left[k^2-m_{\gamma}^2\right]\left[k^2-m_{Z}^2\right]\left[(k-p_{1})^2-m_{l}^2\right]\left[(k-p_{2})^2-m_{l}^2\right]} , \label{ZA}\\
\fl	\Gamma_{\mu}^{(\gamma Z)}=\frac{ge}{2c_{W}}\int\frac{d^4k}{(2\pi)^4}\frac{\gamma_{\alpha}(g^{l}_{V}-g_{A}^{l}\gamma_{5})(\slashed{p_{2}}-\slashed{k}+m_{l})\gamma_{\mu}(\slashed{p_{1}}-\slashed{k}+m_{l})\gamma_{\beta}\, \epsilon^{\alpha\beta\sigma\tau}(k_{AZ})_{\sigma}k_{\tau}}{\left[k^2-m_{\gamma}^2\right]\left[k^2-m_{Z}^2\right]\left[(k-p_{1})^2-m_{l}^2\right]\left[(k-p_{2})^2-m_{l}^2\right]}\: \label{AZ} .
\end{eqnarray}
In the above expressions, $g^l_V=-\frac{1}{2}+2s^2_W$ and $g^l_A=-\frac{1}{2}$. Notice that we have write the amplitude associated with the diagram involving two virtual photons, which disappear in the $(k_{AF})_\mu=0$ approximation. However, it is worth noting that this diagram does not contribute to the MDM, but only to the EDM, since it is proportional to the $\gamma_5$ matrix, as it can be seen after using the identity $\epsilon_{\mu \lambda \rho \alpha }\gamma^\alpha=-i\gamma_5\left(\gamma_\mu \gamma_\lambda \gamma_\rho+g_{\rho \mu}\gamma_\lambda -g_{\mu \lambda }\gamma_\rho -g_{\rho \lambda}\gamma_\mu\right)$. Furthermore, as already mentioned, this diagram induces IR divergences.\\

On the other hand, the amplitudes associated with the diagrams in Fig.~\Ref{Loop1}$(b)$ are given by:
\begin{eqnarray}
	\Gamma_{\mu}^{(WW)(1)}&=&\frac{2}{3}g^2 \int\frac{d^4k}{(2\pi)^4}\frac{P_{R}\: \gamma_{a}\slashed{k}\gamma_{\beta}\, \epsilon_{\alpha\beta\mu\lambda}k_{2}^{\lambda}}{k^2\left[(k-p_{1})^2-m_{W}^2\right]\left[(k-p_{2})^2-m_{W}^2\right]}\: ,\\
	\Gamma_{\mu}^{(WW)(2)}&=&\frac{g^2}{2}\int\frac{d^4k}{(2\pi)^4}\frac{P_{R}\: \gamma^{\beta}\slashed{k}\gamma_{\lambda}\, \epsilon^{\alpha\lambda\sigma\tau}(k_{2})_{\sigma}(k-p_{1})_{\tau}\Gamma_{\alpha\beta\mu}}{k^2\left[(k-p_{1})^2-m_{W}^2\right]^2\left[(k-p_{2})^2-m_{W}^2\right]}\: ,\\
	\Gamma_{\mu}^{(WW)(3)}&=&\frac{g^2}{2}\int\frac{d^4k}{(2\pi)^4}\frac{P_{R}\: \gamma_{\lambda}\slashed{k}\gamma^{\alpha}\, \epsilon^{\lambda\beta\sigma\tau}(k_{2})_{\sigma}(k-p_{2})_{\tau}\Gamma_{\alpha\beta\mu}}{k^2\left[(k-p_{1})^2-m_{W}^2\right]\left[(k-p_{2})^2-m_{W}^2\right]^2}\: \, ,
\end{eqnarray}
where
\begin{eqnarray}
	\Gamma_{\alpha\beta\mu}=g_{\alpha\beta}(-2k_{\mu}+2p_{2\mu}+q_{\mu})-2q_{\beta}g_{\alpha\mu}+2q_{\alpha}g_{\beta\mu}\: ,
\end{eqnarray}
and $P_R=\frac{1}{2}(1+\gamma_5)$ is the right-handed projection operator. Note that all of these amplitudes are proportional to the Lorentz coefficient $(k_2)_\mu$.\\

The contribution to the MDM comes from those terms of the amplitudes that do not involve the matrix $\gamma_5$, whereas the contribution to the EDM arises from terms proportional to this matrix. So, defining $\bar{\Gamma}_{\mu}^{(\gamma Z)}\equiv \Gamma_{\mu}^{(\gamma Z)}+\Gamma_{\mu}^{(Z \gamma )}$, we can write:
\begin{eqnarray}
		\fl\Gamma_{\mu}^{(\gamma\gamma)}=\frac{\alpha}{4\pi}\left[\tilde{f}^{\gamma \gamma}_{1}(q^2) k_{AF}\cdot (p_{1}+p_{2})\: \gamma_{\mu}+\tilde{f}^{\gamma Z}_{2} (q^2)\left(\frac{k_{AF}\cdot q}{m_{l}^2}\right)\frac{(p_{1}+p_{2})_{\mu}}{m_{l}}\right. \nonumber\\
		\fl \qquad\:\:\:\: +\tilde{f}^{\gamma \gamma}_{3}(q^2)\left( \frac{k_{AF}\cdot (p_1+p_2)}{m_Z^2}\right) q_{\mu}+\tilde{f}^{\gamma \gamma}_{4}(q^2) (p_{1}+p_{2})_{\mu} {\pFMSlash{k}_{AF}}+\tilde{f}^{\gamma \gamma}_{5}(q^2)[\gamma_{\mu}, {\pFMSlash{k}_{AF}}]\Bigg]\gamma_5 \, ,\nonumber\\
		{}\\
		\fl\Gamma_{\mu}^{(ZZ)}=\frac{\alpha \:g^{l}_{A}g^{l}_{V}}{16\pi s_W^2 c_W^2}\left[f^{ZZ}_{1}(q^2) k_{ZZ}\cdot (p_{1}+p_{2})\: \gamma_{\mu}+f^{ZZ}_{2} (q^2)\left(\frac{k_{ZZ}\cdot (p_{1}+p_{2})}{m_{Z}^2}\right)\frac{(p_{1}+p_{2})_{\mu}}{2m_{l}}\right. \nonumber\\
		\fl\qquad\:\:\:\: +f^{ZZ}_{3}(q^2)\left( \frac{k_{ZZ}\cdot q}{q^2}\right) q_{\mu}+f^{ZZ}_{4}(q^2)(k_{ZZ})_{\mu}+f^{ZZ}_{5}(q^2) (p_{1}+p_{2})_{\mu} {\pFMSlash{k}_{ZZ}}\Bigg] \nonumber\\
		\fl\qquad \quad	+\frac{\alpha}{16\pi s_W^2 c_W^2}\left[\tilde{f}^{ZZ}_{1}(q^2) k_{ZZ}\cdot (p_{1}+p_{2})\: \gamma_{\mu}+\tilde{f}^{ZZ}_{2} (q^2)\left(\frac{k_{ZZ}\cdot q}{m_{Z}^2}\right)\frac{(p_{1}+p_{2})_{\mu}}{m_{Z}}\right. \nonumber\\
		\fl \qquad\:\:\:\: +\tilde{f}^{ZZ}_{3}(q^2)\left( \frac{k_{ZZ}\cdot (p_1+p_2)}{m_Z^2}\right) q_{\mu}+\tilde{f}^{ZZ}_{4}(q^2) (p_{1}+p_{2})_{\mu} {\pFMSlash{k}_{ZZ}}+\tilde{f}^{ZZ}_{5}(q^2)[\gamma_{\mu}, {\pFMSlash{k}_{ZZ}}]\Bigg]\gamma_5\, ,\nonumber\\
		{}\\
		\fl\bar{\Gamma}_{\mu}^{(\gamma Z)}=\frac{\alpha \:g^{l}_{A}}{8 \pi s_W c_W}\left[f^{\gamma Z}_{1}(q^2) k_{AZ}\cdot (p_{1}+p_{2})\: \gamma_{\mu}+f^{\gamma Z}_{2} (q^2)\left(\frac{k_{AZ}\cdot (p_{1}+p_{2})}{ m_Z^2}\right)\frac{(p_{1}+p_{2})_{\mu}}{2m_{l}}\right. \nonumber\\
		\fl\qquad\:\:\:\: +f^{\gamma Z}_{3}(q^2)\left( \frac{k_{AZ}\cdot q}{q^2}\right) q_{\mu}+f^{\gamma Z}_{4}(q^2)(k_{A Z})_{\mu}+f^{\gamma Z}_{5}(q^2) (p_{1}+p_{2})_{\mu} {\pFMSlash{k}_{AZ}}\Bigg] \nonumber\\
		\fl\qquad \quad	+\frac{\alpha \:g^{l}_{V}}{8 \pi s_W c_W}\left[\tilde{f}^{\gamma Z}_{1}(q^2) k_{AZ}\cdot (p_{1}+p_{2})\: \gamma_{\mu}+\tilde{f}^{\gamma Z}_{2} (q^2)\left(\frac{k_{AZ}\cdot q}{ m_Z^2}\right)\frac{(p_{1}+p_{2})_{\mu}}{m_{Z}}\right. \nonumber\\
		\fl \qquad\:\:\:\: +\tilde{f}^{\gamma Z}_{3}(q^2)\left( \frac{k_{AZ}\cdot (p_1+p_2)}{m_Z^2}\right) q_{\mu}+\tilde{f}^{\gamma Z}_{4}(q^2) (p_{1}+p_{2})_{\mu} {\pFMSlash{k}_{AZ}}+\tilde{f}^{\gamma Z}_{5}(q^2)[\gamma_{\mu}, {\pFMSlash{k}_{AZ}}]\Bigg]\gamma_5 \, ,\nonumber\\
		{}\\
		\fl\Gamma_{\mu}^{(WW)}=\frac{\alpha}{4\pi s_{W}^2}\left[f^{WW}_{1}(q^2) k_{2}\cdot (p_{1}+p_{2})\: \gamma_{\mu}+f^{WW}_{2} (q^2)\left( \frac{k_{2}\cdot (p_{1}+p_{2})}{m_{Z}^2}\right)\frac{(p_{1}+p_{2})_{\mu}}{2m_{l}}\right. \nonumber\\
		\fl\qquad \qquad\:\:\:\: +f^{WW}_{3}(q^2)\left( \frac{k_{2}\cdot q}{q^2}\right) q_{\mu}+f^{WW}_{4}(q^2)(k_{2})_{\mu}+f^{WW}_{5}(q^2) (p_{1}+p_{2})_{\mu} {\pFMSlash{k}_{2}}\Big] \nonumber\\
		\fl\qquad \quad	+\frac{\alpha}{4\pi s_{W}^2}\left[\tilde{f}^{WW}_{1}(q^2) k_{2}\cdot (p_{1}+p_{2})\: \gamma_{\mu}+\tilde{f}^{WW}_{2} (q^2)\left(\frac{k_{2}\cdot q}{m_{Z}^2}\right)\frac{(p_{1}+p_{2})_{\mu}}{m_{Z}}\right. \nonumber\\
		\fl \qquad\:\:\:\: +\tilde{f}^{WW}_{3}(q^2)\left( \frac{k_{2}\cdot (p_1+p_2)}{m_Z^2}\right) q_{\mu}+\tilde{f}^{WW}_{4}(q^2) (p_{1}+p_{2})_{\mu} {\pFMSlash{k}_{2}}+\tilde{f}^{WW}_{5}(q^2)[\gamma_{\mu}, {\pFMSlash{k}_{2}}]\Bigg]\gamma_5\, .\nonumber\\
	\end{eqnarray}
 All the form factors $f^{ZZ}_{i}(q^2)$, $f^{\gamma Z}_{i}(q^2)$, $f^{WW}_{i}(q^2)$, with $i=1,...,5$,
are free of both UV and IR divergences and converge when $q^{2}=0$. In particular, $f^{ZZ}_{3}(0)=f^{\gamma Z}_{3}(0)=f^{WW}_{3}(0)=0$ in this limit. Furthermore, these form factors are absent for an on-shell photon. The form factors which contribute to the MDM are $f_{2}^{XY}(q^2)$ at $q^2=0$, with $XY=\gamma Z,ZZ, WW$. On the other hand, the form factors $\tilde{f}^{XY}_i(q^2)$, with $XY=ZZ,AZ,WW$ are free of both UV and IR divergences. In the case of the $\tilde{f}^{\gamma \gamma}_i(q^2)$ ($i=1,\cdots,5$) form factors, all are free of UV divergences, but $\tilde{f}^{\gamma \gamma}_2(q^2)$, $\tilde{f}^{\gamma \gamma}_3(q^2)$, and $\tilde{f}^{\gamma \gamma}_4(q^2)$ present IR divergences. All form factors are well defined at $q^2=0$. The contributions to the EDM come from the form factors $\tilde{f}^{XY}_2(0)$, with $XY=\gamma \gamma, ZZ, AZ, WW$. In the collinear case ($(k_{AF})_\mu=0$), the $\tilde{f}^{\gamma \gamma}(0)$ is absent, so all contributions to the EDM are free of divergences.\\

After using Gordon's identity and performing some algebraic manipulations, we obtain the contributions of the various diagrams to the MDM. These contributions can be written as:
\begin{eqnarray}
		a_{ZZ}&=&\frac{\alpha \:g^{l}_{A}g^{l}_{V}}{16\pi s_W^2 c_W^2} \left(\frac{k_{ZZ}\cdot(p_{1}+p_{2})}{m_{Z}^2}\right)\mathcal{A}_{ZZ}\: ,\\
		a_{\gamma Z}&=&\frac{\alpha \:g^{l}_{A}}{8 \pi s_W c_W}\left(\frac{k_{AZ}\cdot(p_{1}+p_{2})}{ m_Z^2}\right)\mathcal{A}_{\gamma Z}\: ,\\
		a_{WW}&=&\frac{\alpha}{4\pi s_{W}^2}\left(\frac{k_{2}\cdot(p_{1}+p_{2})}{m_{Z}^2}\right)\mathcal{A}_{WW} \: ,
	\end{eqnarray}
	with $\mathcal{A}_{XY}=f^{XY}_{2}(0)$. So, to first order, the total contribution to the MDM is given by:
\begin{equation}
	\label{al1}
	a^{loop(1)}_l=a_{ZZ}+a_{\gamma Z}+a_{WW}\, .
\end{equation}
The expressions for the form factors $\mathcal{A}_{XY}$ are given in Appendix A.\\

In terms of the LV coefficients $(k_i)_\mu$, the contribution to the MDM given in Eq.~(\Ref{al1}) can be written as follows:
\begin{eqnarray}
		\label{loop1}
		a_{l}^{loop (1)}= \frac{\alpha}{32\pi} \left[ G_{1} \frac{(p_{1}+p_{2})\cdot k_{1}}{m_Z^2} +G_{2} \frac{(p_{1}+p_{2})\cdot k_{2}}{ m_Z^2} \right]\: ,
	\end{eqnarray}
	where
	\begin{eqnarray}
		G_{1}&=&-8g_{A}^{l}\,  \mathcal{A}_{\gamma Z}+\frac{4 g_{A}^{l}g_{V}^{l}}{c_{W}^2} \mathcal{A}_{ZZ}\: ,\\
		G_{2}&=&4 g_{A}^{l}\, \mathcal{A}_{\gamma Z}+\frac{2 g_{A}^{l}g_{V}^{l}}{s_{W}^2} \mathcal{A}_{ZZ}+\frac{8}{s_{W}^2} \mathcal{A}_{WW}\: .	
	\end{eqnarray}
	Using the relation (\ref{cr}), the expression (\ref{loop1}) becomes
	\begin{eqnarray}
		a_{l}^{loop (1)}&=& \frac{\alpha}{32\pi}F^{(1)}\frac{ (p_1+p_2)\cdot k_2}{m_Z^2}\, \\
		&=& \frac{\alpha}{32\pi}\left(-\frac{2c^2_W}{s^2_W}\right)F^{(1)}\frac{(p_1+p_2)\cdot k_1}{ m_Z^2}\, ,
	\end{eqnarray}
	where
	\begin{equation}
		F^{(1)}=\frac{8g^l_Ag^l_Vc_{2W}}{c^2_W s^2_{2W}}{\cal A}_{ZZ}+\frac{4g^l_A}{c^2_W}{\cal A}_{\gamma Z}+\frac{8}{s^2_W}{\cal A}_{WW}\, .
	\end{equation}
\ \\

Regarding the contribution to EDM, it can be written as
	\begin{eqnarray}
		d_{\gamma\gamma}&=&\frac{\alpha}{4\pi} \left(\frac{q\cdot k_{AF}}{m_{l}^2}\right)\left(\frac{e}{m_l}\right)\tilde{\mathcal{A}}_{\gamma\gamma}\: ,\\
		d_{ZZ}&=&\frac{\alpha}{16\pi s_W^2 c_W^2} \left(\frac{q\cdot k_{ZZ}}{m_{Z}^2}\right)\left(\frac{e}{m_Z}\right)\tilde{\mathcal{A}}_{ZZ}\: ,\\
		d_{\gamma Z}&=&\frac{\alpha \:g^{l}_{V}}{8 \pi s_W c_W}\left(\frac{q\cdot k_{AZ}}{ m_Z^2}\right)\left(\frac{e}{m_Z}\right)\tilde{\mathcal{A}}_{\gamma Z}\: ,\\
		d_{WW}&=&\frac{\alpha}{4\pi s_{W}^2}\left(\frac{q\cdot k_{2}}{m_{Z}^2}\right)\left(\frac{e}{m_Z}\right)\tilde{\mathcal{A}}_{WW} \: ,
	\end{eqnarray}
	with $\tilde{\mathcal{A}}_{XY}=\tilde{f}^{XY}_{2}(0)$. Although we are working on the $(k_{AF})_\mu=0$ approximation, we have included the expression for $d_{\gamma \gamma}$ for completeness. So, to first order in the LV coefficients, the total contribution to the EDM is given by
	\begin{eqnarray}\label{dl1}
		d_l^{loop(1)}=d_{ZZ}+d_{\gamma Z}+d_{WW}\, .
	\end{eqnarray}
 The expressions for the form factors $\tilde{\mathcal{A}}_{XY}$ are given in Appendix A. Analogous with the MDM, the expression (\ref{dl1}) can be written in terms of the LV coefficients $(k_i)_\mu$ as follows:
	\begin{eqnarray}\label{loop1dE}
		d_{l}^{loop (1)}= \frac{\alpha}{32\pi}\left(\frac{e}{m_Z}\right)\left[\tilde{G}_{1} \, \frac{q\cdot k_{1}}{m_Z^2} +\tilde{G}_{2}\, \frac{q\cdot k_{2} }{ m_Z^2} \right]\: ,
	\end{eqnarray}
	where
	\begin{eqnarray}
		\tilde{G}_{1}&=&-8g_{V}^{l}\,  \tilde{\mathcal{A}}_{\gamma Z}+\frac{4}{c_{W}^2}\tilde{\mathcal{A}}_{ZZ}\: ,\\
		\tilde{G}_{2}&=&4g_{V}^{l}\, \tilde{\mathcal{A}}_{\gamma Z}+\frac{2}{s_{W}^2} \tilde{\mathcal{A}}_{ZZ}+\frac{8}{s_{W}^2} \tilde{\mathcal{A}}_{WW}\: .	
	\end{eqnarray}
	In the $(k_{AF})_\mu=0$ approximation, using the relation (\ref{cr}), the expression (\ref{loop1dE}) becomes:
	\begin{eqnarray}
\label{Ft1}
		d_{l}^{loop (1)}&=& \frac{\alpha}{32\pi}\left(\frac{e}{m_Z}\right)\tilde{F}^{(1)}\left(\frac{q\cdot k_2}{m_Z^2}\right)\, \\
		&=& \frac{\alpha}{32\pi}\left(-\frac{2c^2_W}{s^2_W}\right)\left(\frac{e}{m_Z}\right)\tilde{F}^{(1)}\left( \frac{q\cdot k_1}{m_Z^2}\right)\, ,
	\end{eqnarray}
	where
	\begin{equation}
		\tilde{F}^{(1)}=\frac{8c_{2W}}{c^2_W s^2_{2W}}\tilde{\mathcal{A}}_{ZZ}+\frac{4g^l_V}{c^2_W}\tilde{\mathcal{A}}_{\gamma Z}+\frac{8}{s^2_W}\tilde{\mathcal{A}}_{WW}\, .
	\end{equation}

\subsubsection{$O(k^2)$ contribution}\label{L2}
\ \\
\noindent To second order in the LV coefficients $(k_i)_\mu$, the contributions to the electromagnetic properties of the charged lepton $l$ are given by the diagrams shown in Fig.\Ref{Loop2}. We will focus only on the energy-independent contributions. In this case, we have verified that there are no contributions to the EDM. The CPT-Odd contribution to the vertex function can be organized as follows:
\begin{eqnarray}
	\Gamma^{O(2)}_\mu=\Gamma^{(\gamma \gamma \gamma)}_\mu+\Gamma^{(ZZZ)}_\mu+\Gamma^{(\gamma \gamma Z)}_\mu+\Gamma^{(Z\gamma \gamma )}_\mu+\Gamma^{(\gamma ZZ)}_\mu+\Gamma^{(ZZ\gamma)}_\mu\nonumber\\
	\qquad \quad+\Gamma^{(\gamma Z\gamma)}_\mu+\Gamma^{(Z\gamma Z)}_\mu+\Gamma^{(WWW)}_\mu \, ,
\end{eqnarray}
where the labels $(\gamma \gamma \gamma)$, $\cdots$, indicate the number of gauge boson propagators in diagrams of Fig.~\Ref{Loop2}$(a)$. The $\Gamma^{(WWW)}_\mu$ amplitude represents the sum of amplitudes associated with diagrams of Fig.~\Ref{Loop2}$(b)$.\\

In conventional field theories, the most general electromagnetic $\bar{f}f\gamma$ coupling, with $f$ a charged fermion, is given by:
\begin{equation}
	\label{FV}
	ie\, \Gamma_\mu=ie\left[F_1(q^2)\gamma_\mu +F_2(q^2)\frac{i\sigma_{\mu \nu}q^\nu}{2m_f}-F_3(q^2)\sigma_{\mu \nu}q^\nu \gamma_5\right]\, ,
\end{equation}
where $q$ is the moment of the photon. With the photon on-shell, the form factors $F_i$ represent the following physical quantities: $F_1(0)=Q_f$ is the charge carried by $f$, $F_2(0)=a_f$ is the MDM of $f$, and $-Q_fF_3(0)=d_f$ is the electric dipole moment of $f$. At the one-loop level, the on-shell $\Gamma_\mu$ vertex is proportional to $\gamma_\mu$ and to the moment of the incoming fermion, $p_{1\mu}$. The Gordon's identity allows us to express $p_{1\mu}$ in terms of the Lorentz structures given in (\ref{FV}), that is, $\frac{p_{1\mu}}{m_f}=\gamma_\mu+\frac{i\sigma_{\mu \nu}q^\nu}{2m_f}$. So, in our case, we first focus on the coefficients $k_{i}\cdot k_{j}$ of the diverse amplitudes, with $k_{i,j}=k_{AF},k_{ZZ},k_{AZ},k_2$. Next, we look in these expressions for the coefficient of $\frac{p_{1\mu}}{m_l}$, which corresponds to the contribution to the MDM $a_l$. Contributions to $a_l$ that depend on energy can, inprinciple, arise, but we will focus only in the usual on-shell contribution $F_2(0)=a_l$. We organize the CPT-Odd contribution to $a_l$ given by the diverse diagrams of Figs.~\ref{Loop2}$(a)$ and \ref{Loop2}$(b)$ as follows:
\begin{eqnarray}
\fl a^{O(2)}_l=a^{\gamma \gamma \gamma }_l+a^{ZZZ}_l+a^{\gamma \gamma Z }_l+a^{Z\gamma \gamma }_l+a^{Z Z \gamma }_l+a^{\gamma Z Z }_l +a^{\gamma Z \gamma }_l+a^{Z \gamma Z}_l+a^{W W W }_l\, .
\end{eqnarray}
As already commented, this diagram-by-diagram contribution of Fig.~\ref{Loop2}$(a)$ is possible because the corresponding amplitudes are UV-divergences-free and gauge-independent by themselves. As we will see below, IR divergences only are present in the $a^{\gamma \gamma \gamma }_l$ contribution, but this contribution does not exist in the $(k_{AF})_\mu=0$ approximation.\\

From the above considerations, the amplitudes for the diagrams of Fig.~\ref{Loop2}$(a)$ can be written as follows:
\begin{equation}
	\label{ppp}
	\Gamma^{(\gamma \gamma \gamma)}_\mu = ie^2\int \frac{d^4k}{(2\pi)^4} \frac{\left[\gamma_\beta(\pFMSlash{k}+\pFMSlash{p_2}+m_l)\gamma_\mu(\pFMSlash{k}+\pFMSlash{p_1}+m_l)\gamma_\alpha\right] T^{\alpha \beta}_{(AF)(AF)}}{[k^2-m^2_\gamma]^3[(k+p_1)^2-m^2_l][(k+p_2)^2-m^2_l]}\, ,
\end{equation}
\begin{eqnarray}
	\label{zzz}
\fl	\Gamma^{(ZZZ)}_\mu =\frac{ig^2}{4c^2_W}\int \frac{d^4k}{(2\pi)^4} \frac{\left[\gamma_\beta(g^l_V-g^l_A\gamma_5)(\pFMSlash{k}+\pFMSlash{p_2}+m_l)\gamma_\mu(\pFMSlash{k}+\pFMSlash{p_1}+m_l)\gamma_\alpha (g^l_V-g^l_A\gamma_5)\right] }{[k^2-m^2_Z]^3[(k+p_1)^2-m^2_l][(k+p_2)^2-m^2_l]}\nonumber\\
\qquad \times T^{\alpha \beta}_{(ZZ)(ZZ)}\, ,
\end{eqnarray}

\begin{equation}
	\label{ppz}
\fl	\Gamma^{(\gamma \gamma Z)}_\mu = \frac{ieg}{2c_W}\int \frac{d^4k}{(2\pi)^4} \frac{\left[\gamma_\beta(g^l_V-g^l_A\gamma_5)(\pFMSlash{k}+\pFMSlash{p_2}+m_l)\gamma_\mu(\pFMSlash{k}+\pFMSlash{p_1}+m_l)\gamma_\alpha \right]T^{\alpha \beta}_{(AF)(AZ)}}{[k^2]^2[k^2-m^2_Z][(k+p_1)^2-m^2_l][(k+p_2)^2-m^2_l]}\, ,
\end{equation}

\begin{equation}
	\label{zpp}
\fl	\Gamma^{(Z \gamma \gamma )}_\mu =\frac{ieg}{2c_W}\int \frac{d^4k}{(2\pi)^4} \frac{\left[\gamma_\beta(\pFMSlash{k}+\pFMSlash{p_2}+m_l)\gamma_\mu(\pFMSlash{k}+\pFMSlash{p_1}+m_l)\gamma_\alpha (g^l_V-g^l_A\gamma_5)\right]T^{\alpha \beta}_{(AZ)(AF)}}{[k^2]^2[k^2-m^2_Z][(k+p_1)^2-m^2_l][(k+p_2)^2-m^2_l]}\, ,
\end{equation}

\begin{equation}
	\label{pzz}
\fl	\Gamma^{(\gamma Z Z)}_\mu = \frac{ieg}{2c_W}\int \frac{d^4k}{(2\pi)^4} \frac{\left[\gamma_\beta(g^l_V-g^l_A\gamma_5)(\pFMSlash{k}+\pFMSlash{p_2}+m_l)\gamma_\mu(\pFMSlash{k}+\pFMSlash{p_1}+m_l)\gamma_\alpha \right]T^{\alpha \beta}_{(AZ)(ZZ)} }{[k^2][k^2-m^2_Z]^2[(k+p_1)^2-m^2_l][(k+p_2)^2-m^2_l]}\, ,
\end{equation}

\begin{equation}
	\label{zzp}
\fl	\Gamma^{(ZZ \gamma)}_\mu = \frac{ieg}{2c_W}\int \frac{d^4k}{(2\pi)^4} \frac{\left[\gamma_\beta(\pFMSlash{k}+\pFMSlash{p_2}+m_l)\gamma_\mu(\pFMSlash{k}+\pFMSlash{p_1}+m_l)\gamma_\alpha (g^l_V-g^l_A\gamma_5)\right]T^{\alpha \beta}_{(ZZ)(AZ)} }{[k^2][k^2-m^2_Z]^2[(k+p_1)^2-m^2_l][(k+p_2)^2-m^2_l]}\, ,
\end{equation}

\begin{equation}
	\label{pzp}
	\Gamma^{(\gamma Z \gamma)}_\mu = ie^2\int \frac{d^4k}{(2\pi)^4} \frac{\left[\gamma_\beta(\pFMSlash{k}+\pFMSlash{p_2}+m_l)\gamma_\mu(\pFMSlash{k}+\pFMSlash{p_1}+m_l)\gamma_\alpha \right]T^{\alpha \beta}_{(AZ)(AZ)} }{[k^2]^2[k^2-m^2_Z][(k+p_1)^2-m^2_l][(k+p_2)^2-m^2_l]}\, ,
\end{equation}

\begin{eqnarray}
\fl	\label{zpz}
	\Gamma^{(Z \gamma Z)}_\mu = \frac{ig^2}{4c^2_W}\int \frac{d^4k}{(2\pi)^4} \frac{\left[\gamma_\beta(g^l_V-g^l_A\gamma_5)(\pFMSlash{k}+\pFMSlash{p_2}+m_l)\gamma_\mu(\pFMSlash{k}+\pFMSlash{p_1}+m_l)\gamma_\alpha  (g^l_V-g^l_A\gamma_5)\right] }{[k^2][k^2-m^2_Z]^2[(k+p_1)^2-m^2_l][(k+p_2)^2-m^2_l]}\nonumber\\
\qquad	\times T^{\alpha \beta}_{(AZ)(AZ)}\, .
\end{eqnarray}
In the above expressions,
\begin{eqnarray}
	T^{\alpha \beta}_{(X)(Y)}&=&\left[ \epsilon^{\lambda \alpha \sigma \tau } (k_{X})_\sigma k_\tau \right]g_{\lambda \rho} \left[ \epsilon^{\rho \beta \eta \zeta } (k_{Y})_\eta k_\zeta \right]\nonumber \\
	&=&\left[(k_X \cdot k)(k_Y\cdot k)-(k_X\cdot k_Y)k^2 \right]g^{\alpha \beta}+k^\alpha_Yk^\beta_X k^2\nonumber \\
	&&+(k_X\cdot k_Y)k^\alpha k^\beta -(k_X\cdot k)k^\alpha_Y k^\beta -(k_Y\cdot k)k^\beta_X k^\alpha \, ,
\end{eqnarray}
where $X,Y=AF,AZ,ZZ$. Note that we have introduced a fictitious mass for the photon only in diagram $(\gamma \gamma \gamma)$, since it is the only one with IR divergences. \\

Regarding the contribution of the diagrams in Fig.~\ref{Loop2}$(b)$, it can be divided into two parts as follows:
\begin{equation}
	\Gamma^{WWW}_\mu= \Gamma^{(1+2)}_\mu+ \Gamma^{(3+4+5)}_\mu \, ,
\end{equation}
where $\Gamma^{(1+2)}_\mu$ represents the contributions of the first and second diagrams, which are characterized by the presence of the CPT-Odd $WW\gamma$ vertex. This contributions is given by:
\begin{eqnarray}
	\Gamma^{(1+2)}_\mu=-\frac{2ig^2}{3}\int \frac{d^4k}{(2\pi)^4}\frac{\epsilon_{\alpha \mu \lambda \eta}k^\eta_2}{k^2[(k-p_1)^2-m^2_W][(k-p_2)^2-m^2_W]}P_R\nonumber\\
	\qquad\qquad\times\left[\gamma_\rho \pFMSlash{k} \gamma^\lambda \frac{\epsilon^{\alpha \rho \sigma \tau }k_{2\sigma } (k-p_2)_\tau}{(k-p_2)^2-m^2_W} + \gamma^\lambda \pFMSlash{k} \gamma_\rho \frac{\epsilon^{\alpha \rho \sigma \tau }k_{2\sigma } (k-p_1)_\tau}{(k-p_1)^2-m^2_W}\right]\, .
\end{eqnarray}
On the other hand, $\Gamma^{(3+4+5)}_\mu$ represents the contribution of the last three diagrams in Fig.~\ref{Loop2}$(b)$. These diagrams are characterized by the contribution of the SM $WW\gamma$ vertex. This amplitude is given by:
\begin{eqnarray}
	\Gamma^{(3+4+5)}_\mu=ig^2\int \frac{d^4k}{(2\pi)^4} \frac{\Gamma_{\lambda \rho \mu}}{k^2[(k-p_1)^2-m^2_W][(k-p_2)^2-m^2_W]}P_R\nonumber\\
	\qquad\qquad\qquad\times \Bigg\{\gamma_\alpha \pFMSlash{k} \gamma_\beta\frac{ \epsilon^{\alpha \lambda \sigma \tau}k_{2\sigma}(k-p_2)_\tau  \epsilon^{\beta \rho \sigma \tau}k_{2\sigma}(k-p_1)_\tau}{[(k-p_1)^2-m^2_W][(k-p_2)^2-m^2_W]} \nonumber \\
	\qquad\qquad\qquad+\gamma^\lambda \pFMSlash{k} \gamma_\sigma\frac{\epsilon^{\alpha \rho \tau \eta}k_{2\tau}(k-p_1)_\eta \,  g_{\alpha \beta} \, \epsilon^{\beta \sigma \tau \eta}k_{2\tau}(k-p_1)_\eta }{[(k-p_1)^2-m^2_W]^2}\nonumber\\
	\qquad\qquad\qquad+\gamma_\sigma \pFMSlash{k} \gamma^\rho \frac{\epsilon^{\alpha \sigma \tau \eta}k_{2\tau}(k-p_2)_\eta \,  g_{\alpha \beta} \, \epsilon^{\beta \lambda \tau \eta}k_{2\tau}(k-p_2)_\eta  }{[(k-p_2)^2-m^2_W]^2} \Bigg\}\, ,
\end{eqnarray}
where $k_{2\lambda}$ is the LV coefficient. In addition,
\begin{equation}
	\Gamma_{\lambda \rho \mu}=q_\rho g_{\lambda \mu}-q_\lambda g_{\rho \mu}-(k-p_1)_\mu g_{\lambda \rho}\, ,
\end{equation}
is the SM $WW\gamma$ vertex function. \\

Using the FeynCalc package~\cite{FCP}, the contributions of diagrams in Fig.~\ref{Loop2}$(a)$ to the MDM can be expressed as follows:
\begin{eqnarray}
\label{a21}
	a^{\gamma \gamma \gamma}_l&=&\frac{\alpha}{32\pi} \left(\frac{k^2_{AF}}{m^2_l}\right){\cal A}_{\gamma \gamma \gamma}\, ,\\
\label{a22}
	a^{ZZZ}_l&=&\frac{\alpha}{32\pi} \left(\frac{k^2_{ZZ}}{m^2_l}\right){\cal A}_{ZZZ}\, \\
\label{a23}
a^{\gamma \gamma Z}_l&=&a^{Z\gamma \gamma}_l=\frac{\alpha}{32\pi}\frac{1}{s_{2W}}\left(\frac{k_{AF}\cdot k_{ZZ}}{m^2_l}\right)g^l_V{\cal A}_{\gamma \gamma Z}\, ,\\
\label{a24}
a^{\gamma Z Z}_l&=&a^{Z Z \gamma}_l=\frac{\alpha}{32\pi}\frac{1}{s_{2W}}\left(\frac{k_{AZ}\cdot k_{ZZ}}{m^2_l}\right)g^l_V{\cal A}_{\gamma ZZ}\, ,\\
\label{a25}
a^{\gamma Z \gamma}_l&=&\frac{\alpha}{32\pi}\left(\frac{k^2_{AZ}}{m^2_l}\right){\cal A}_{\gamma Z \gamma}\, ,\\
\label{a26}
a^{Z\gamma Z}_l&=&\frac{\alpha}{32\pi}\frac{1}{s^2_{2W}}\left(\frac{k^2_{AZ}}{m^2_l}\right){\cal A}_{Z\gamma Z}\, .
\end{eqnarray}
On the other hand, the contributions of diagrams in Fig.~\ref{Loop2}$(b)$ to the MDM are given by:

\begin{eqnarray}
\label{a27}
	a^{(1+2)}_l&=&\frac{\alpha}{32\pi s^2_W}\left(\frac{k^2_2}{m^2_l}\right){\cal A}_{(1+2)}\, ,\\
\label{a28}
	a^{(3+4+5)}_l&=&\frac{\alpha}{32\pi s^2_W}\left(\frac{k^2_2}{m^2_l}\right){\cal A}_{(3+4+5)}\, ,
\end{eqnarray}
The various form factors appearing in Eqs(\ref{a21}-\ref{a28}) are given in Appendix B. Note that the amplitudes ${\cal A}_{\gamma \gamma \gamma}$ and ${\cal A}_{\gamma \gamma Z}$ are absent in the $(k_{AF})_\mu=0$ limit, but we have included them for completeness.\\

It is convenient to write the total CPT-Odd contribution to $O(k^2)$ in terms of the $k_{1\mu}$ and $k_{2\mu}$ Lorentz coefficients. After using Eqs.~(\ref{CLAA}), (\ref{CLZZ}), and (\ref{CLAZ}), we obtain
\begin{eqnarray}
	\label{AMDM}
	a^{loop(2)}_l&=&\frac{\alpha}{32\pi}\left[F_1\left(\frac{k^2_1}{m^2_l}\right)+F_{12}\left(\frac{k_1\cdot k_2}{m^2_l}\right)+F_2\left(\frac{k^2_2}{m^2_l}\right) \right]\, ,
\end{eqnarray}
where
\begin{eqnarray}
\fl F_1=4\left(c^4_W{\cal A}_{\gamma \gamma \gamma}+s^4_W{\cal A}_{ZZZ}\right)-2g^l_V\left(c^2_W {\cal A}_{\gamma \gamma Z}+s^2_W{\cal A}_{\gamma ZZ}\right)+s^2_{2W}{\cal A}_{\gamma Z \gamma}+{\cal A}_{Z\gamma Z}\, , \\
\fl	F_{12}=s^2_{2W}({\cal A}_{\gamma \gamma \gamma}+{\cal A}_{ZZZ})+c_{2W}g^l_V({\cal A}_{\gamma \gamma Z}-{\cal A}_{\gamma ZZ})-{\cal A}_{Z\gamma Z}-s^2_{2W}{\cal A}_{\gamma Z \gamma} \, , \\
\fl F_2=s^4_W{\cal A}_{\gamma \gamma \gamma}+c^4_W{\cal A}_{ZZZ}+\frac{1}{2}g^l_V\left(s^2_W{\cal A}_{\gamma \gamma Z}+c^2_W{\cal A}_{\gamma ZZ}\right)+\frac{1}{4}\left(s^2_{2W}{\cal A}_{\gamma Z \gamma}+{\cal A}_{Z\gamma Z}\right)\nonumber\\
\fl \qquad\, +\frac{1}{s^2_W}{\cal A}_{WWW}\, .
\end{eqnarray}
In these expressions, $s_{2W}=2s_Wc_W$ and $c_{2W}=c^2_W-s^2_W$. In the collinear case, we use the relation (\ref{cr}) to obtain
\begin{eqnarray}
	\label{loop2}
	a^{loop(2)}_l&=&\frac{\alpha}{32\pi}\, F^{(2)}\frac{k^2_2}{m^2_l}\, \\
	&=&\frac{\alpha}{32\pi}\,\left(-\frac{2c^2_W}{s^2_W}\right) F^{(2)}\frac{k^2_1}{m^2_l}\, ,
\end{eqnarray}
where
\begin{eqnarray}
	F^{(2)}=\frac{c^2_{2W}}{c^4_W}{\cal A}_{ZZZ}+\left(g^l_V \frac{c_{2W}(4-s^2_{2W})}{8c^4_W}+\frac{s^6_W}{c^2_W}\right){\cal A}_{\gamma ZZ}\nonumber\\
	\qquad+\frac{s^2_W}{4c^2_W}\left(2g^l_Vc_{2W}+\frac{s^2_W}{c^2_W}+\frac{c^2_W}{s^2_W}\right){\cal A}_{\gamma Z \gamma}+\frac{1}{s^2_W}{\cal A}_{WWW}\, .
\end{eqnarray}

\section{Discussion}
\label{Dis}We now proceed to discuss our results. We will start by analyzing the contribution to the EDM. As seen in the previous section, the EDM is generated at the one-loop level. Since the contribution is linear in the LV coefficients, the EDM depends on the energy, in this case of the transferred momentum $q_\mu$. In  Table~\ref{TLEDM} we show the various loop amplitudes defining the contributions to the EDM of the three lepton families. From this Table, we can see that the contribution to the EDM is essentially determined by the diagram characterized by a photon and a Z boson, whose associated amplitude is $g^l_V \tilde{\mathcal{A}}_{\gamma Z}$. As shown in Appendix A, this strong effect associated with a light lepton arises from the fact that the $\tilde{\mathcal{A}}_{\gamma Z}$ amplitude behaves like $\frac{m_Z}{m_l}$, while the  $\tilde{\mathcal{A}}_{ZZ}$ and $\tilde{\mathcal{A}}_{WW}$ amplitudes behave like $\frac{m_l}{m_Z}$. In the case of the electron EDM, it can be appreciated from Table~\ref{TLEDM} that the amplitude $g^l_V \tilde{\mathcal{A}}_{\gamma Z}$ is approximately nine orders of magnitude larger than the $\tilde{\mathcal{A}}_{ZZ}$ and $\tilde{\mathcal{A}}_{WW}$ amplitudes.
\begin{table}[h]
	\caption{Numerical values of the various form factors that define the CPT-Odd contribution to the electric moment of charged leptons to $O(k)$.} \label{TLEDM}
	\begin{indented}
		\lineup
		\scriptsize
		\item[]\begin{tabular}{@{}*{7}{@{\extracolsep{12pt plus 0pt}}l}}
			\br
			$m_{l}$  & $\tilde{\mathcal{A}}_{ZZ}$ & $g_{V}^{l}\tilde{\mathcal{A}}_{\gamma Z}$ & $\tilde{\mathcal{A}}_{WW}$ & $\tilde{G}_{1}$& $\tilde{G}_{2}$ & $\tilde{F}^{(1)}$\\
			\mr
			$m_{e}$  & $8.16 \times 10^{-7}$     &  $-1.07\times 10^{3}$       &  $4.51 \times 10^{-7}$     &  $8.55\times 10^{3}$    &   $-4.27\times 10^{3}$ & $-5.50\times 10^{3}$ \\
			$m_{\mu}$ & $1.69 \times 10^{-4}$     &  $-5.17$ &  $9.33 \times 10^{-5}$     &  $4.14\times 10^{1}$    &   $-2.07\times 10^{1}$ & $-2.66\times 10^{1}$\\
			$m_{\tau}$ & $2.80 \times 10^{-3}$     &   $-3.07\times 10^{-1}$     &  $1.57 \times 10^{-3}$     &  $2.47$    &  $-1.15$ & $-1.50$\\
			\br
		\end{tabular}
	\end{indented}
\end{table}

Let $d_l^{exp}$ be the experimental limit on the EDM of the lepton $l$. To establish bounds on the LV coefficients $(k_i)_\mu$, we demand that the CPT-Odd contribution be less that the experimental limit. From Eq.(\ref{Ft1}) we obtain the following condition
\begin{eqnarray}\label{deCond1}
	\left|\frac{q\cdot k_2}{m_Z^2}\right|<\frac{32\pi}{\alpha e}\frac{|d_l^{exp}|}{|\tilde{F}^{(1)}|} m_Z \, .
\end{eqnarray}
As already commented, this condition depends on the photon momentum $q$. Since the EDM is defined on-shell, i.e. at $q^2=0$, which implies $E_\gamma\equiv q_0=|\mathbf{q}|$, we obtain
\begin{eqnarray}\label{qk21}
	(q\cdot k_2)=E_\gamma\left[(k_2)_0-|\mathbf{k}_2|\cos\theta_\gamma\right]\, ,
\end{eqnarray}
with $\theta_\gamma$ the angle between the 3-momentum of the emitted photon and the spatial part of $(k_2)_\mu$.

On the other hand, the most stringent experimental limit on an EDM of a charged lepton is that of the electron. However, to set a limit on the LV coefficient $(k_2)_\mu$ through EDM, we must know the energy of the photons emitted in the experimental device. The ACME collaboration found an upper limit of $|d_e|<1.1\times 10^{-29}e\,cm$ \cite{ACME} at 90 \% CL. This collaboration obtains this limit by measuring the spin precession of electrons subjected to intramolecular electric fields of thorium monoxide. In this experiment, the Collaboration detects the spontaneous emission of photons with a wavelength of 512 nm. In terms of the $Z$ boson mass $m_Z$, the emitted photons have an energy of $E_\gamma=2.64\times 10^{-8}m_Z$. From these considerations, we obtain the following bound
\begin{equation}
\label{EDM1}
 |(k_2)_0-|\mathbf{k_2}|cos\theta_\gamma |<0.862\, m_e \, .
\end{equation}

An interesting case corresponds to a light-like $(k_2)_\mu$ vector. In this special case, (\ref{EDM1}) takes the form 
\begin{equation}
\label{EDM2}
 |(k_2)_0(1-cos\theta_\gamma) |<0.862\, m_e \, .
\end{equation}

We now move on to the analysis of the contributions to the MDM. The total CPT-Odd contribution to the MDM of a charged lepton $l$ up to the one loop level and up to second order in the LV coefficients $(k_i)_\mu$ is given by by the sum of Eqs.~(\ref{tree1}), (\ref{tree2}), (\ref{loop1}), and (\ref{loop2}):
\begin{equation}
	\label{at}
	a^{CPT-Odd}_l=a^{tree(1)}_l+a^{tree(2)}_l +a^{loop(1)}_l +a^{loop(2)}_l \, .
\end{equation}
These contributions can be classified into two categories, namely those that are energy dependent, $a^{tree(1)}_l$ and $a^{loop(1)}_l$, and those that are energy independent  , $a^{tree(2)}_l$ and $a^{loop(2)}_l$. In Tables \ref{TL1} and \ref{TL2}, we show the relative importance of the diverse quantities defining the CPT-Odd contribution to the MDM of charged leptons at the one loop level to first and second order in the LV coefficients, respectively.

We will first study the energy-dependent contributions to the MDM. From Table~\ref{TL1} we can see that, as it occurs in the EDM case, the one-loop contribution is determined essentially by the diagram with a photon and a $Z$ boson circulating in the loop. In the case of the electron, the amplitude ${\cal A}_{\gamma Z}$ is eleven and twelve orders of magnitude larger than the amplitudes ${\cal A}_{ZZ}$ and ${\cal A}_{WW}$, respectively. In Appendix A, we show that for $\frac{m_l}{m_Z}\ll 1$, ${\cal A}_{\gamma Z}\sim O(1)$, whereas  ${\cal A}_{ZZ}\sim \left(\frac{m_l}{m_Z}\right)^2$ and ${\cal A}_{WW}\sim 10^{-1}\times \left(\frac{m_l}{m_Z}\right)^2$. On the other hand, from equation (\ref{tree1}) and table~\ref{TL1}, it can be seen that the momentum-dependent contribution is dominated by the tree-level effect. In fact, for any of the three charged leptons, we find that $\frac{a_l^{tree(1)}}{a_l^{loop(1)}}\approx 10^3$. Because of this, we will focus on the tree-level contribution. We now proceed to estimate a bound for the LV coefficient $(k_2)_\mu$ from this contribution to the MDM. To do this we will use data from a recently reported measurement~\cite{MDMe} of the electron MDM. The experimental device is known as a Penning trap, which  essentially consists of a cylindrical arrangement with a uniform magnetic field pointing in the positive z direction, $\vec{B}=B\, \hat{e_z}$, and an electrostatic quadrupole potential, $V\propto z^2-\frac{\rho^2}{2}$, with $\rho^2=x^2+y^2$. The magnetic moment operator associated with the electron is related with its spin operator $\vec{S}$, normalized to its spin eigenvalue $\frac{\hbar}{2}$, as follows:
\begin{equation}
	\vec{\mu}=-\frac{g}{2}\mu_B \left(\frac{\vec{S}}{\frac{\hbar}{2}}\right)\, ,
\end{equation}
where $\mu_B=\frac{e\hbar}{2m_e}$ is the Bohr magneton and $g$ is the Land\' e factor. The energy levels of the trapped electron are given by
\begin{equation}
	E_{nm_s}=h\nu_s m_s+h\nu_c\left(n+\frac{1}{2}\right)\, , \, \, \, \, \, m_s=\pm \frac{1}{2}\, , \, \, \, \, n=0,1,2,\cdots ,
\end{equation}
with $h=2\pi \hbar$, $\nu_c=\frac{eB}{2\pi m_e}$ the cyclotron frequency, and $\nu_s=(\frac{g}{2}) \nu_c$ the spin frecquency. Since the trapped electron rotates in the $xy$ plane, the average of the linear momentum components in this plane can be taken equal to zero, that is, $\bar{p}_x$ and $\bar{p}_y$. If we consider, in addition $p_z=0$, the energy-dependent factor $F^{(1)}\frac{(p_1+p_2)\cdot k_2}{m^2_e}$ becomes
\begin{equation}
	\frac{(p_1+p_2)\cdot k_2}{4m^2_W}=\left(\frac{E}{2m_W}\right)\left(\frac{(k_2)_0}{m_W}\right)\, .
\end{equation}
Experimentally, it is considered an energy $E=6.78\times 10^{-9} \, GeV$, which corresponds to the eigenvalues $n=10$ and $m_s=+\frac{1}{2}$.

Demanding that this $O(k_i)$ tree-level contribution be less than the experimental uncertainty $\Delta a^{\scriptsize\emph{Exp}}_e=\pm 2.8\times 10^{-13}$~\cite{PDG} on the electron MDM,
\begin{equation}
\left| \frac{(p_1+p_2)\cdot k_2}{4m_W^2}\right| <|\Delta a_e^{Exp}|\, ,
\end{equation}
we obtain the following bounds for the time components of the LV coefficients $(k_i)_\mu$:
\begin{eqnarray}
		|(k_2)_0| <2.09 \times 10^{3} m_e \, ,\\
		|(k_1)_0| <3.00 \times 10^{2} m_e \, .
\end{eqnarray}
We can see that this bound is three orders of magnitude less stringent than that obtained in Eq.(\ref{EDM1}) from the experimental limit on the electron EDM. Consequently, we can conclude that the best bound for linear effects of the LV coefficients is obtained from the experimental limit on EDM. So, from now on, we  will focus only on $O(k^2)$ effects on the electron MDM.
\begin{table}[h]
	\caption{Numerical values of the various form factors that define the CPT-Odd contribution to the magnetic moment of charged leptons to $O(k)$.} \label{TL1}
	\begin{indented}
		\lineup
		\scriptsize
		\item[]\begin{tabular}{@{}*{7}{@{\extracolsep{12pt plus 0pt}}l}}
			\br
			$m_{l}$  & $\mathcal{A}_{ZZ}$ & $g_{A}^{l}\mathcal{A}_{\gamma Z}$ & $\mathcal{A}_{WW}$ & $G_{1}$& $G_{2}$ & $F^{(1)}$\\
			\mr
			$m_{e}$  & $1.05 \times 10^{-11}$     &  $6.67\times 10^{-1}$       &  $-2.17 \times 10^{-12}$     &  $-5.33$    &   $2.67$ & $3.43$ \\
			$m_{\mu}$ & $4.47 \times 10^{-7}$     &  $6.67\times 10^{-1}$       &  $-9.26 \times 10^{-8}$     &  $-5.33$    &   $2.67$ & $3.43$  \\
			$m_{\tau}$ & $1.26 \times 10^{-4}$     &  $6.67\times 10^{-1}$       &  $-2.62 \times 10^{-5}$     &  $-5.33$    &   $2.67$ & $3.43$ \\
			\br
		\end{tabular}
	\end{indented}
\end{table}

We will now proceed to study the contributions of $O(k^2)$ to the MDM of the charged lepton $l$. First, notice that the $\frac{a^{tree(2)}_l}{a^{loop(2)}_l}$ ratio is of the order of $10^{-8}$, $10^{-3}$, and $10^{-1}$ for the $e$, $\mu$, and $\tau$ leptons, respectively. This result shows us that the tree-level contribution can be ignored compared to the one-loop contribution. The reason why a loop-level contribution may be more important than a tree-level contribution will be clarified later. 

So, neglecting the tree level contributions and the energy-dependent one-loop contribution, the contribution to the MDM of the lepton $l$ becomes
\begin{equation}
	a^{CPT-Odd}_l=\frac{\alpha}{32\pi}F^{(2)}\frac{k^2_2}{m^2_l}\, .
\end{equation}
\begin{table}[h]
	\caption{Numerical values of the various form factors that define the CPT-Odd contribution to the magnetic moment of charged leptons to $O(k^2)$. \label{TL2}}
	\lineup
	\scriptsize
	\begin{tabular*}{490pt}{@{}l*{11}{@{\extracolsep{7pt plus 20pt}}l}}
		\br
		$m_{l}$  & $\m{\cal A}_{ZZZ}$ &  $\m g^l_V\, {\cal A}_{\gamma ZZ}$  &  $\m {\cal A}_{\gamma Z \gamma}$ &  $\m {\cal A}_{Z\gamma Z}$ & $\m{\cal A}_{WWW}$ & $\m F_1$ & $\m F_{12}$ & $\m F_2$ &  $\m F^{(2)}$\\
		\mr
		$m_{e}$  & $-1.05\times 10^{1}$ &  $2.69\times 10^{-2}$ & $-5.00\times 10^{-1}$ & $-1.26\times 10^{-1}$ & $ -0.755208$ & $-2.61$ & $-6.79$ &$-9.82$ & $-8.84$\\
		$m_\mu$ &  $-5.43$ & $2.69\times 10^{-2}$ &  $-5.00\times 10^{-1}$ &$-1.26\times 10^{-1}$ & $-0.755209$ & $-1.61$ & $-3.29$ & $-6.77$ & $-6.27$\\
		$m_\tau$ & $-2.79$& $2.69\times 10^{-2}$  &  $-4.42\times 10^{-1}$ &$-1.27\times 10^{-1}$ & $-0.755523$ & $-1.04$ & $-1.50$ & $-5.17$ & $-4.92$ \\
		\br		
	\end{tabular*}
\end{table}
To begin, we would like to make some comments about the behavior of one-loop amplitudes. First, notice that all one-loop amplitudes have a common factor of $\frac{k^2_2}{m^2_l}$. Later we will address the reason why the mass $m_l$ of the lepton $l$ and not the mass $m_Z$ of the $Z$ boson should appear in this factor. The various loop functions ${\cal A}_{i}$ have interesting characteristics that deserve careful study. Examined as functions of the variable $x=\frac{m_l}{m_Z}$, we can see from Appendix B that, with the exception of the amplitude ${\cal A}_{ZZZ}$ , the rest of all Amplitudes tend to a constant non-zero value in the $x\to 0$ limit. In this limit, the amplitude ${\cal A}_{ZZZ}$ diverges logarithmically. Due to the presence of the global factor $\frac{k^2_2}{m^2_l}$, the limit $x\to 0$ makes sense only for $m_l$ fixed and $m_Z\to \infty$. This behavior of the amplitudes shows us a strong non-decoupling effect. The non-decoupling effect is even more pronounced in the case of the amplitude ${\cal A}_{ZZZ}$, which grows as $\log\left(\frac{m_Z}{m_l}\right)$. In fact, as we can see in Table~\ref{TL2}, the contribution to the MDM is essentially determined by the amplitude ${\cal A}_{ZZZ}$. From these considerations it follows that contribution to the MDM of the electron is by far the most important. From now on we will focus on the MDM of the electron.

To impose a bound on $k^2_2$, we demand that the CPT-Odd contribution be less than  experimental uncertainty $\Delta a^{\scriptsize\emph{Exp}}_e$. So, we have:

\begin{equation}
	\left|F^{(2)}\,\frac{ k^2_2}{m^2_e} \right|< \frac{32\pi}{\alpha}\left|\Delta a^{\scriptsize\emph{Exp}}_e \right|\, .
\end{equation}
Using the experimental data, we obtain the following bound
\begin{equation}
\label{k2b}
\left|\frac{ k^2_2}{m^2_e} \right|<4.36\times 10^{-10}\, .
\end{equation}

On the other hand, we have emphasizing throughout the paper that the $(k_i)_\mu$ ($i=1,2$) CPT-Odd coefficients have positive mass dimension. As a consequence of this fact, we can talk about the existence of the real scalar $k^2_2$ (remember that we are using the collinear approximation given by Eq.(\ref{cr})), which is invariant under OLT and has squared mass dimension. This fact naturally implies the existence of a new physics scale if it is assumed that $k^2_2>0$. The scale given by  $k^2_2=\Lambda^2_{CPT}$, which will be the same for all inertial observers, would characterize the CPT violation phenomena. So from Eq.(\ref{k2b}) the following bound for the $\Lambda_{CPT}$ scale is obtained:
\begin{equation}
	\label{B15}
	\Lambda_{CPT}< 2.08\times 10^{-5} m_e\, .
\end{equation}
It should be noted that this result constitutes an upper limit, in contrast to those derived from the physical effects associated with new heavy particles, which correspond to  lower limits. We now proceed to clarify this point and other unusual aspects already noted above.

Our results obtained above seem to go against certain well-established technical aspects of perturbation theory in conventional Quantum Field Theory (QFT). The result that may seem most surprising is that we have obtained an upper bound for the new-physics scale $\Lambda_{CPT}$, when we have learned the opposite from conventional theories. Furthermore, from the perspective of perturbation theory, it does not seem natural that a tree-level contribution is  suppressed relative to a one-loop contribution. We now proceed to clarify these points.

In conventional QFTs that are renormalizable by the power counting criterion, any parameter with positive mass units is always associated with the mass of a field. If such a mass is generated by the Higgs mechanism, it may also appear through some vertices of the theory. The point to keep in mind is that when the considered field contributes virtually to a given physical process, its mass appears through the corresponding propagator. When conventional theories beyond the SM are considered, they incorporate new fields that are heavy in the sense that they have masses proportional to a scale well above the Fermi scale. In physical processes in which new heavy fields only contribute virtually, the propagators of such fields play a central role, since their contributions tend to disappear when their masses are much larger than the Fermi scale. In that case, we say that the new physics decouples~\cite{DeThe}. In this context, when we use precision electroweak data, lower limits can be obtained on the masses of the heavy fields. Lower bounds on energy scales can also be derived in the context of effective field theories that include interactions of dimension greater than four, as it occurs in the most general case of the SME. For dimensional considerations, these types of interactions are suppressed by appropriate inverse powers of a new-physics scale. As mentioned in the introduction, this type of parameters, which are characterized by having negative mass units, belong to the category of irrelevant parameters and are always associated with non-renormalizable interactions of dimension greater than four. By considering such a contribution to a SM observable, a lower bound on the new-physics scale can be obtained. In our case, the LV coefficients $(k_i)_\mu$ have positive mass units which means that, as mentioned in the introduction, they belong to the category of relevant parameters, which are associated with interactions of dimension less than four. Thus, it seems natural to obtain upper bounds for this type of parameters.

The mSME is not an extension of the SM in the sense of a conventional renormalizable theory such as the one commented above. This despite the fact that the mSME is a renormalizable theory by the power counting criterion. First of all, is the fact that this theory does not incorporate new degrees of freedom, since it is made only from SM fields. Secondly, as mentioned in the introduction, the parameters of the theory are dimensionless (CPT-Even) or have a positive unit of mass (CPT-Odd). The key point here is that the parameters with units of mass, as in the case at hand, do not correspond to the masses of some fields, since there are no new fields, they simply quantify the intensity of some couplings among SM fields. This is the reason why the contribution of these parameters does not appear through propagators, but only through vertices. This fact explains why the new physics effects do not decouple in the $(k_i)_\mu \to \infty$ (or $k^2_i \to \infty$) limit, but rather in the $(k_i)_\mu \to 0$. The fact that decoupling occurs for small LV coefficients is implicit in our assumption of considering the corresponding couplings as a perturbation. All this is in agreement with the upper bound that we have obtained for the $\Lambda_{CPT}$ scale.

Regarding the unusual result that tree-level contributions are suppressed relative to the one-loop level contributions, it has to do with the fact that the LV coefficients $(k_i)_\mu$ have mass units which gives rise to a non-decoupling effect of a heavy field circulating in the loops. This heavy particle is the $Z$ gauge boson, whose mass is several orders of magnitude larger than the lepton masses. This phenomenon arises at second order in the LV coefficients. As we have seen, the tree-level contribution to the MDM (remember that there is no second order contribution to the EDM) is determined by the exchange of a virtual Z boson, which leads to a contribution proportional to $\frac{k^2_i}{m^2_Z}$. At the one-loop level, to illustrates the idea, we analyze two cases characterized by the first and fourth diagrams in Fig.\Ref{Loop2}. Notice that the first diagram is entirely determined by the QEDE (this diagram has IR divergences, but we will concentrate on the finite part, as it is sufficient for our purposes). As it was shown above, these diagrams are gauge-independent and free of UV divergences by themselves, so physical conclusions can be derived from each of them. In the first diagram of this figure, the particles circulating in the loop are the photon and the charged lepton, so the only natural energy scales that determine the contribution to the MDM are the LV coefficient $(k_i)_\mu$ and the mass of the lepton. For dimensional considerations and due to the fact that the LV coefficients only appear through  vertices, such a contribution must be proportional to $\frac{k^2_i}{m^2_l}$. Regarding the fourth diagram, the particles that circulate in the loop are the photon, the $Z$ boson, and the charged lepton $l$. In this case, there are three natural energy scales, namely $(k_i)_\mu$, $m_Z$, and $m_l$. From dimensional considerations, two types of possible contributions to the MDM can be thought of, one having the form $\left(\frac{k^2_i}{m^2_l}\right)f(x)$ and the other having the form $\left(\frac{k^2_i}{m^2_Z}\right) g (x)$, with $x=\frac{m_l}{m_Z}$. Physically, $f$ or $g$ must be well-behaved functions on $x$. Our results in Appendix B show that $f(x)\to cte\neq 0$ when $x\to 0$, but $g(x)=\frac{f(x)}{x^2}$ clearly diverges in this limit. The fact that all form factors ${\cal A}_i$ characterizing the second order one-loop contribution tend to a non-zero value in the $x\to 0$ limit means that the heavy-physics effects (the $Z$ gauge boson, whose mass is much greater than the mass $m_l$ of the process) do not decouple. This non-decoupling effect is even more pronounced in the case of the form factor ${\cal A}_{ZZZ}$, as it tends to infinity logarithmically in the $x\to 0$ limit (see Appendix B). These results show us that the one-loop level contribution  distinguishes the flavor of the charged lepton, while the contribution at the tree-level does not. This selection between lepton species is quite pronounced, as $\frac{k^2_i}{m^2_\mu}\approx 2.5 \times 10^{-5} \, \frac{k^2_i}{m^2_e}$ and $\frac{k^2_i}{m^2_\tau}\approx  10^{-7} \,  \frac{k^2_i}{m^2_e}$, this in contrast with the tree-level contribution which does not depend on the lepton mass. These results also show us that, as emphasized in the introduction, these new-physics effects are highly sensitive to low-energy processes, since the energies involved in measuring the MDM of the electron are of the order of the mass of this particle. To contrast the high sensitivity of the LV coefficients $(k_i)_\mu$ to low-energy processes, the contribution to the MDM of the electron should be compared with the corresponding contribution to the MDM of the top quark (high-energy system). In this case, the factors $\frac{k^2_i}{m^2_e} $ and $\frac{k^2_i}{m^2_t} $ differ by eleven orders of magnitude. These results become even more important if we take into account that the electron MDM is measured with a much higher degree of experimental precision than that of the rest of the leptons. This is one of the best scenarios to constraint physics beyond SM.

We would like to emphasize that this non-decoupling effect arises at the one-loop level and up to the second order in the LV coefficients $(k_i)_\mu$. This contrasts with the one-loop linear contributions, which are of decoupling nature, as deduced from the corresponding expressions for the EDM and MDM. In this case, such contributions disappear in the $m_Z \to \infty$ limit.

\section{Conclusions}
\label{Con}In this paper, the impact of the CPT-Odd electroweak gauge sector of the mSME on the electromagnetic properties of charged leptons was studied. This sector is characterized by relevant interactions in the sense that they have canonical dimension less than four. An interesting consequence of these types of interactions is that they have very strong effects on low energy observables. We have calculated the contributions of the dimension-three operators characterizing this sector to the electromagnetic properties of the three charged leptons, up to second order in the LV coefficients $(k_i)_\mu$, both at the tree and one-loop levels. We find that contributions of $O(k_i)$ depend on the energy, whereas $O(k^2_i)$ contributions are independent on energy. It is found that contributions at the tree level do not distinguish between lepton species, but those generated at the one-loop level do. The MDM of the lepton $l$ is found to receive contributions from both $O(k_i)$ and $O(k^2_i$), the one and the other generated at both the tree and one-loop levels. However, it is found that the EDM of these leptons is only induced at the one-loop level and at $O(k_i)$, making it energy dependent. We have taken advantage of the fact that the Lorentz coefficient $(k_{AF})_\mu$ of the gauge sector of QEDE is given by a linear combination of the Lorentz vectors $(k_1)_\mu$ and $(k_2)_\mu$, the first associated with the gauge group $U_Y(1)$ and the second with the gauge group $SU_L(2)$, and is strongly constrained by the experiment in order to establish a link between $(k_1 ) _ \mu$ and $(k_2 )_\mu$. Motivated by this, we have assumed that $(k_{AF})_\mu$ is equal to zero. This implies that the vectors $(k_1)_\mu$ and $(k_2)_\mu$ are collinear, which leads to establishing limits linked to each other. It is found that, at order $O(k_i)$, the one-loop contribution to the MDM is approximately three orders of magnitude less important than the tree-level contribution. Under certain reasonable assumptions about the experimental setup relative to the tree-level energy-dependent contribution, we have been able to obtain bounds on the LV coefficients. Using recent experimental results on the electron MDM the tree-level contribution leads to bounds on the time component of the LV coefficients $(k_i)_\mu$ of $|(k_2)_0|< 2.09 \times 10^3\, m_e$ and $|(k_1)_0|< 3 \times 10^2\, m_e$. A more restrictive bound of
$|(k_2)_0-|\mathbf{k_2}|\cos\theta_\gamma|<0.86\, m_e$ was obtained using a recent experimental limit on the electron EDM. It is worth mentioning that all these effects of $O(k_i)$ disappear in the $m_Z\gg m_l$ limit.

Regarding the effects of $O(k^2_i)$, there are contributions to the MDM at both the tree and one-loop levels, but there are no contributions to the EDM. The tree-level contribution is determined by the interchange of a $Z$ gauge boson and disappears in the $m^2_Z \gg k^2_i$ limit. In contrast, the one-loop contribution involves the three mass scales of the problem, namely the mass of the lepton in consideration, $m_l$, the mass of the $Z$ gauge boson, $m_Z$, and the LV coefficient $(k_i)_\mu$. In this case, the contribution to the MDM is of the form $\frac{k^2_i}{m^2_l}f\left(\frac{m_l}{m_Z}\right)$, with $f$ a well-behaved function on the $m_l$ and $m_Z$ masses. In this case, it is crucial to investigate the behavior of the $f$ function with respect to the mass ratio $\frac{m_l}{m_Z}$, given the enormous difference that exists between the values of these masses, especially in the case of the electron. We have carried out a careful study of the various one-loop amplitudes that give rise to this function $f\left(\frac{m_l}{m_Z}\right)$. It was found that the amplitudes of all diagrams containing combinations of the photon and the $Z$ gauge boson do not disappear in the  $m_Z \gg m_l$ limit, but rather tend to a constant non-zero value. This means that the heavy-physics effects, characterized by the virtual contribution of the $Z$ gauge boson, do not decouple from the low-energy process, characterized by the mass of the charged lepton $l$. An extreme case occurs with the amplitude associated with the diagram characterized by three $Z$ gauge bosons circulating in the loop, since its amplitude grows logarithmically with $m_Z$ mass. This strong non-decoupling effect causes the one-loop contribution to dominate over the tree-level contribution by a factor of $\left(\frac{m_Z}{m_l}\right)^2$, which, in the case of the electron, is about $10^{10}$. Behind this result is the fact that the LV coefficients $(k_i)_\mu$ belong to the category of relevant parameters in the sense that they have positive units of mass, which means that they are associated with interactions of dimension lower than four. In our case, the $(k_i)_\mu$ paramters are associated with dimension-three interactions. We see that the one-loop contribution can be several orders of magnitude larger than the tree-level contribution, but we must keep in mind the fact that the former is of $O(k^2_i)$ while the latter is of $O(k_i ) $ and that we have treated these effects as a perturbation, which means that the LV coefficients are assumed to be small. In marked contrast to conventional QFT, these parameters with dimension of a positive mass unit are not associated with masses of new fields, which means that they do not contribute to a given process through propagators, which would lead to a compensatory effect, but simply quantify interactions among  SM fields.

Using the experimental data on the electron MDM, we have obtained the following upper bound for $k^2_2$:
\begin{equation}
	\left|\frac{ k^2_2}{m^2_e} \right|<4.36\times 10^{-10}\, .
\end{equation}
Motivated by the fact that the four-vector $(k_2)_\mu$ has mass units, we have introduced the new physics scale $\Lambda_{CPT}=\sqrt{k^2_2}$ and obtained the upper bound $\Lambda_{CPT}<2.08\times 10^{-5}\, m_e$.  We would like to emphasize that this bound is invariant under observer Lorentz transformations.

Schwonger's calculation of the QED correction of one-loop, $\frac{\alpha}{2\pi}$, to the MDM receives corrections in the SM proportional to $\frac{m^2_l}{m_V^2}$ and $\frac{ m^2_l}{m_H^2}$ of the weak gauge bosons ($V=W,Z$) and the Higgs boson, respectively. In conventional extensions of the SM characterized by a new physical scale $M$, these  corrections are proportional to positive powers of the mass ratio $\frac{m_l}{M}$. The main characteristic of all these corrections is that they decouple in the limit when the mass of the new particles is much greater than the mass of the lepton under consideration. In this work, we have presented one-loop level results that incorporate second-order effects on the LV coefficients, which present novel characteristics that deviate significantly from this behavior. We think that our results could motivate the study of radiative corrections in the context of the SME beyond the first order in the LV coefficients.
 
\section*{Appendix A: Loop amplitudes of $O(k)$ contribution}
\label{AA}
The MDM form factors $\mathcal{A}_{XY}$ are given by:
\begin{eqnarray}
\fl\mathcal{A}_{ZZ}=\left(\frac{m_Z}{m_l}\right)^2\left(-2+\frac{3}{2}\frac{m_Z^2}{m_l^2}\right)+\frac{1}{2}\left(\frac{m_Z}{m_l}\right)^4 \left[ 5B_0(1)+6B_0(2)-11B_0(3)\right]\nonumber\\
\fl\qquad \quad +\left(\frac{m_Z}{m_l}\right)^2 \left[-5 B_0(1)-2B_0(2)+7B_0(3)\right]+\left(6-12\frac{m_Z^2}{m_l^2}+\frac{15}{4}\frac{m_Z^4}{m_l^4}\right)m_Z^2 C_0(1)\nonumber\\
\fl\qquad \quad +\left(-2+7\frac{m_Z^2}{m_l^2}-\frac{11}{4}\frac{m_Z^4}{m_l^4}\right)m_Z^2 C_0(2)+\frac{1}{4}\left(24-24\frac{m_Z^2}{m_l^2}+5\frac{m_Z^4}{m_l^4}\right)m_Z^4 D_0(1)\, ,
\end{eqnarray}
\begin{eqnarray}
\fl\mathcal{A}_{\gamma Z}=\left(\frac{m_Z}{m_l}\right)^2\left(4-4\frac{m_l^2}{m_Z^2}-\frac{3}{2}\frac{m_Z^2}{m_l^2}\right)+\left(\frac{m_Z}{m_l}\right)^2\left[5B_0(1)+2B_0(2)-7B_0(3)\right]\nonumber \\
\fl \qquad\quad +2\left[B_0(3)-B_0(1)\right]-\frac{1}{4}\left(\frac{m_Z}{m_l}\right)^4\left[5B_0(1)+6B_0(2)-11B_0(3)\right]\nonumber\\
\fl \qquad\quad +\left(6-6\frac{m_l^2}{m_Z^2}-\frac{5}{4}\frac{m_Z^2}{m_l^2}\right) m_Z^2 C_0(1)\, ,
\end{eqnarray}
\begin{eqnarray}
\fl\mathcal{A}_{WW}=\frac{1}{48c_W^2}\left(\frac{m_W}{m_l}\right)^2\left(7+9\frac{m_W^2}{m_l^2}\right)+\frac{1}{16c_W^2}\left(\frac{m_W}{m_l}\right)^4\left(11-6\frac{m_l^2}{m_W^2}\right)\left[B_0^W(1)-B_0^W(2)\right]\nonumber\\
\fl\qquad\quad +\frac{1}{16c_W^2} \left[\left(1-\frac{22}{3}\frac{m_W^2}{m_l^2}+5\frac{m_W^4}{m_l^4}\right)C_0^W(3)-\left(2-15\frac{m_W^2}{m_l^2}+13\frac{m_W^4}{m_l^4}\right) C_0^W(4)\right]m_W^2\nonumber\\
\fl\qquad\quad -\frac{1}{16c_W^2} \left(\frac{m_W}{m_l}\right)^2 \left(1+5\frac{m_W^2}{m_l^2}\right)(m_l^2-m_W^2)^2 D_0^W(1)\, .
\end{eqnarray}
In the above expressions, the factors $B_0(i)$, $C_0(i)$, and $D_0(i)$ are Passarino-Veltman (PV) scalar functions of two, three, and four points, respectively, which are given in Appendix C. In obtaining these results, the FeynCalc package \cite{FCP} was used. All PV functions appearing in both $O(k)$ and $O(k^2)$ have relatively simple solutions in terms of elementary functions, which are given in Appendix C. In terms of the variables $x=\frac{m_l}{m_z}$ and $y=\frac{m_l}{m_W}$, we can express the form factors $\mathcal{A}_{XY}$ with some few terms of the corresponding power series in $x$ or $y$ with a good approximation to the analytical solutions because $x,y\ll 1$. The first terms of such power series are:

\begin{eqnarray}\label{AzzxPower}
	\mathcal{A}_{ZZ}=\frac{1}{3}x^2-\frac{8}{5}x^4+O(x^6)\, ,
\end{eqnarray}
\begin{eqnarray}\label{AazxPower}
	\mathcal{A}_{\gamma Z}=-\frac{4}{3}+\frac{x^2}{3}+O(x^4)\, \, ,
\end{eqnarray}
\begin{eqnarray}\label{AwwxPower}
	\mathcal{A}_{WW}=-\frac{1}{120 c_W^2}\left(5y^2+4y^4\right)+O(y^6)\, .
\end{eqnarray}
Note that the form factors $\mathcal{A}_{ZZ}(x)$and $\mathcal{A}_{WW}(y)$ tend to zero in the $x, y \to 0$ limits, but $\mathcal{A}_{\gamma Z}(x)\to -\frac{4}{3}$ in this limit. These limits are evident in the series power expansion, but we have also verified them using the complete analytical solutions. Note that the contributions of the form factors $\mathcal{A}_{ZZ}(x)$ and $\mathcal{A}_{WW}(y)$ are marginal with respect to that of $\mathcal{A}_{\gamma Z}(x)$.\\

On another hand, the EDM form factors $\tilde{A}_{XY}$, are given by:
\begin{eqnarray}\label{Aaatilde}
	\tilde{\mathcal{A}}_{\gamma\gamma}=\tilde{f}_{\gamma \gamma}+\lim_{q^2\rightarrow 0}\frac{\tilde{g}_{\gamma \gamma}(q^2)}{q^2}\, ,
\end{eqnarray}
where
\begin{eqnarray}
	\fl  \tilde{f}_{\gamma \gamma} =-\frac{7}{12}+\frac{1}{24}\left(13B^{\gamma}_0(3)-12B^\gamma_0(2)-B_0(1)\right)+\frac{7}{12}\left(B^\gamma_0(2)-B^\gamma_0(3)\right)-m_l^2C^\gamma_0(1)\, ,
\end{eqnarray}
\begin{eqnarray}
	\fl  \tilde{g}_{\gamma \gamma}(q^2) =\frac{4m_l^4}{(q^2-4m_l^2)^2}\Bigg\{-2m_\gamma^2 +2m_l^2\left(B_0(1)-2B_0(1,q^2)+2B^\gamma_0(2)-B^\gamma_0(3)\right)\\
	\fl \qquad\qquad +2m_\gamma^2\left(B_0(1,q^2)-2B^\gamma_0(2)+B^\gamma_0(3)\right)+(m_\gamma^2-2 m_l^2)m_\gamma^2C^\gamma_0(2)\nonumber\\
	\fl \qquad\qquad +(3m_\gamma^4-12 m_\gamma^2 m_l^2+8m_l^4)C^\gamma_0(1,q^2)+\left(m_\gamma^4-6m_\gamma^2m_l^4+8 m_l^4 \right)m_\gamma^2 D^\gamma_0(1,q^2)\Bigg\} .
\end{eqnarray}
The scalar functions $B^\gamma_0(i)$, $B^\gamma_0(i,q^2)$, $C^\gamma_0(i)$, $C^\gamma_0(i,q^2)$ and $D^\gamma_0(i,q^2)$ are the PV functions $B_0(i)$, $B_0(i,q^2)$, $C_0(i)$, $C_0(i,q^2)$ and $D_0(i,q^2)$ of the Appendix C, with the replacement $m_Z\rightarrow m_\gamma$, respectively. In addition $\tilde{g}_{\gamma \gamma}(q^2=0)=0$, so the limit expressed in (\ref{Aaatilde}) must be evaluated carefully.
\begin{eqnarray}
\fl	\tilde{\mathcal{A}}_{ZZ}=\left((g_A^l)^2+(g_V^l)^2\right)^2\left(\tilde{f}_{ZZ}^{+}+\lim_{q^2\rightarrow 0}\frac{\tilde{g}_{ZZ}^{+}(q^2)}{q^2}\right)+\left((g_A^l)^2-(g_V^l)^2\right)^2\tilde{f}_{ZZ}^{-}\,,
\end{eqnarray}
where the factors $\tilde{f}_{ZZ}^{\pm}$ depend on the lepton and $Z$ masses. In addition, $g_{ZZ}^{+}(0)=0$. These expressions are given by:
\begin{eqnarray}
\fl	\tilde{f}_{ZZ}^{+}=-\frac{1}{12}\left(\frac{m_Z}{m_l}\right)^3\left(1-\frac{7}{2}\frac{m_Z^2}{m_l^2}\right)-\frac{1}{12}\left(\frac{m_Z}{m_l}\right)^3\left(5B_0(1)+9B_0(2)-14B_0(3)\right)\nonumber\\
\fl \qquad  \quad +\frac{7}{12}\left(\frac{m_Z}{m_l}\right)^5\left(B_0(2)-B_0(3)\right)-\frac{1}{2}\left(1-\frac{7}{3}\frac{m_Z^2}{m_l^2}+\frac{7}{12}\frac{m_Z^4}{m_l^4}\right)\frac{m_Z^3}{m_l}C_0(2)\, ,
\end{eqnarray}
\begin{eqnarray}
	\fl	\tilde{g}_{ZZ}^{+}(q^2)=\frac{1}{(q^2-4m_l^2)^2}\Bigg\{-8m_l m_z^5 +8m_l m_Z^3\Big[ m_l^2\left(B_0(1)+2B_0(2)-B_0(3)-2B_0(1,q^2)\right)\nonumber\\
	\fl  \qquad\quad \quad\quad +m_Z^2\left(-2B_0(2)+B_0(3)+B_0(1,q^2)\right)\Big]+4m_lm_Z^5(m_Z^2-2m_l^2)C_0(2)\nonumber\\
	\fl  \qquad\quad \quad\quad +4m_lm_Z^3\left(8m_l^4-12m_l^2m_Z^2+3m_Z^4\right)C_0(1,q^2)\nonumber\\
	\fl  \qquad\quad \quad\quad +4m_lm_Z^5\left(8 m_l^4-6m_l^2m_Z^2+m_Z^4\right)D_0(1,q^2)\Bigg\}\, ,
\end{eqnarray}
\begin{eqnarray}
	\fl	\tilde{f}_{ZZ}^{-}=-\frac{1}{8}\left(\frac{m_Z}{m_l}\right)^3\left(3B_0(1)+2B_0(2)-5B_0(3)\right)+\left(1-\frac{3}{4}\frac{m_Z^2}{m_l^2}\right)\frac{m_Z^3}{m_l}C_0(1)\nonumber\\
\fl\qquad\quad-\left(\frac{1}{2}-\frac{5}{8}\frac{m_Z^2}{m_l^2}\right)\frac{m_Z^3}{m_l}C_0(2)+\left(1-\frac{3}{8}\frac{m_Z^2}{m_l^2}\right)\frac{m_Z^5}{m_l}D_0(1)\, .
\end{eqnarray}
\begin{eqnarray}
		\tilde{\mathcal{A}}_{\gamma Z}=\left(\frac{m_Z}{m_l}\right)\left[\tilde{f}_{\gamma Z}+\lim_{q^2\rightarrow 0}\frac{\tilde{g}_{\gamma Z}(q^2)}{q^2}\right]\, ,
\end{eqnarray}
with
\begin{eqnarray}
\fl	\tilde{f}_{\gamma Z}=\frac{7}{12} \left(\frac{m_Z}{m_l}\right)^2\left(2-\frac{m_Z^2}{m_l^2}\right)+\frac{1}{12} \left(\frac{m_Z}{m_l}\right)^2\Bigg[B_0(1)+12B_0(2)-13B_0(3) \nonumber\\
\fl\qquad\qquad\quad +7\frac{m_Z^2}{m_l^2}\left(B_0(3)-B_0(2)\right)\Bigg]+\left(2-\frac{3}{4}\frac{m_Z^2}{m_l^2}\right)m_Z^2 C_0(1)\, ,
\end{eqnarray}
\begin{eqnarray}
	\fl	\tilde{g}_{\gamma Z}(q^2)=\frac{8m_l^2 m_Z^2}{(q^2-4m_l^2)^2}\Bigg\{2m_Z^2-4m_l^2+m_l^2\left(4B_0(1,q^2)-4B_0(2)+2B_0(3)-2B_0(1)\right)\nonumber\\
\fl \qquad\quad\qquad +m_Z^2\left(2B_0(2)-B_0(3)-B_0(1,q^2)\right)-(8m_l^4-6m_l^2 m_Z^2+m_Z^4)C_0(1,q^2)\Bigg\}\, .
\end{eqnarray}
\begin{eqnarray}
	\tilde{\mathcal{A}}_{WW}=\frac{1}{c_W^3}\left(\tilde{f}_{WW}+\lim_{q^2\rightarrow 0}\frac{\tilde{g}_{WW}(q^2)}{q^2}\right)\, ,
\end{eqnarray}
where
\begin{eqnarray}
\fl \tilde{f}_{WW}=\frac{1}{32}\left(\frac{m_W}{m_l}\right)^3\left(1+\frac{7}{3}\frac{m_W^2}{m_l^2}\right)+\frac{7}{48}\frac{m_W^3}{m_l^5}(m_W^2-m_l^2)(B_0^{W}(1)-B^W_0(2))\nonumber\\
\fl \qquad\quad-\frac{1}{48}\left(11-13\frac{m_l^2}{m_W^2}\right)m_W^2 C^W_0(3)+\frac{1}{48}\left(19-\frac{31}{2}\frac{m_l^2}{m_W^2}-\frac{7}{2}\frac{m_W^2}{m_l^2}\right)m_W^2 C^W_0(4)\nonumber\\
\fl \qquad\quad+\frac{1}{4}\left(\frac{m_W}{m_l}\right)^3(m_l^2-m_W^2)^2\, D^W_0(1)\, ,
\end{eqnarray}
\begin{eqnarray}
\fl	\tilde{g}_{WW}(q^2)=\frac{2}{3}\frac{m_l m_W^3}{(q^2-4m_l^2)^2}\Bigg\{-m_l^2 -3m_W^2+3m_W^2\left(B^W_0(1,q^2)+B^W_0(2)-2B^W_0(1)\right)\nonumber\\
\fl \qquad\qquad\quad +5m_l^2\left(B^W_0(1,q^2)-B^W_0(1)\right)+\frac{3}{2}\left(m_W^4+4m_l^2m_W^2-5m_l^4\right)C^W_0(4)\nonumber\\
\fl \qquad\qquad\quad +\left(3m_W^4-2m_l^2m_W^2+3m_l^4\right)C^W_0(3,q^2)+\frac{3}{2}(m_l^2-m_W^2)(5m_l^2+3m_W^3)C^W_0(4,q^2)\nonumber\\
\fl \qquad\qquad\quad -3(m_l^2-m_W^2)^2(m_l^2+m_W^2)D^W_0(1,q^2)\, .
\end{eqnarray}
To evaluate the limits $\lim_{q^2\rightarrow 0}\frac{\tilde{g}_{XY}(q^2)}{q^2}$, we have used the PackageX program \cite{PX}. As in the case of the magnetic form factors, eqs. (\ref{AzzxPower})-(\ref{AwwxPower}), we express the $\tilde{A}_{XY}$ factors as a power series in the $x$ and $y$ variables:
\begin{eqnarray}
	\tilde{\mathcal{A}}_{\gamma\gamma}=\frac{1}{6}\left[4+\log\left(\frac{m_\gamma^2}{m_l^2}\right)\right]\, ,
\end{eqnarray}
\begin{eqnarray}
	\tilde{\mathcal{A}}_{ZZ}=\left((g_A^l)^2+(g_V^l)^2\right)^2\left[\frac{x}{4}+x^3\left(\frac{8}{3}\log x+\frac{149}{45}\right)\right]\nonumber\\
	\quad \quad\:\:\:\: +\left((g_A^l)^2-(g_V^l)^2\right)^2\left[\frac{x}{3}+x^3\left(4\log x+\frac{31}{6}\right)\right]+O(x^5)\, ,
\end{eqnarray}
\begin{eqnarray}
	\tilde{\mathcal{A}}_{\gamma Z}=\frac{1}{9x}-\frac{x}{6}+O(x^3)\, ,
\end{eqnarray}
\begin{eqnarray}
	\tilde{\mathcal{A}}_{WW}=\frac{1}{720 c_W^3}\left(35 y+16 y^3\right)+O(y^5)\, .
\end{eqnarray}
Note that the form factor $\tilde{\mathcal{A}}_{\gamma\gamma}$ is IR divergent. We can see that, as in the case of magnetic form factors, the contribution $\tilde{\mathcal{A}}_{\gamma Z}$ dominates. In fact, while the amplitudes $\tilde{\mathcal{A}}_{ZZ}$ and $\tilde{\mathcal{A}}_{WW}$ tend to zero in the $x,y\to 0$ limit, the amplitude $\tilde{\mathcal{A}}_{\gamma Z}$ diverges in this limit. We have verified these limits using the complete analytical solutions of the amplitudes.

\section*{Appendix B: Loop amplitudes of $O(k^2)$ contribution}
\label{AB}
The various form factors appearing in Eqs.(\ref{a21}-\ref{a28}) are given by:
\begin{eqnarray}
	{\cal A}_{\gamma \gamma \gamma}=\frac{3}{2}+3B_0(5)-2B_0(4)-B_0(1)+4m^2_lC_0(4)-2m^2_\gamma C_0(6)\nonumber\\
	\qquad\qquad+8m^2_lm^2_\gamma D_0(3)+4m^2_lm^4_\gamma E_0(2)\, .
\end{eqnarray}

\begin{equation}
	{\cal A}_{ZZZ}=\left( (g^l_A)^2+(g^l_V)^2 \right)\left(\frac{m_Z}{m_l}\right)^2{\cal A}^+_{ZZZ}+\left( (g^l_A)^2-(g^l_V)^2 \right){\cal A}^-_{ZZZ}\, ,
\end{equation}
with
\begin{eqnarray}
	{\cal A}^+_{ZZZ}=-\frac{3}{16}\left(1-\frac{m^2_l}{m^2_Z}\right)+\frac{3}{32}\left[11B_0(3)-6B_0(2)-5B_0(1) \right]\nonumber\\ \qquad\quad+\frac{1}{8}\frac{m^2_l}{m^2_Z}\left[38B_0(3)-20B_0(2)-18B_0(1) \right]\nonumber\\
	\qquad\quad-\frac{1}{16}\left(15+12\frac{m^2_l}{m^2_Z}-32\frac{m^4_l}{m^4_Z} \right)m^2_Z C_0(1)\nonumber\\
	\qquad\quad+\frac{1}{2}\left(\frac{33}{16}+\frac{19}{6}\frac{m^2_l}{m^2_Z}-\frac{7}{3}\frac{m^4_l}{m^4_Z}\right)m^2_ZC_0(2)-\frac{1}{48}\left(9+20\frac{m^2_l}{m^2_Z} \right)m^2_ZC_0(3)\nonumber \\
	\qquad\quad-\frac{1}{8}\left(5+6\frac{m^2_l}{m^2_Z}-32\frac{m^4_l}{m^4_Z}+16\frac{m^6_l}{m^6_Z} \right)m^4_ZD_0(1)\nonumber\\
	\qquad\quad+\frac{1}{96}\left(33+76\frac{m^2_l}{m^2_Z}-112\frac{m^4_l}{m^4_Z} \right)m^4_ZD_0(2)\nonumber \\
	\qquad\quad-\frac{1}{32}\left(5+8\frac{m^2_l}{m^2_Z}-64\frac{m^4_l}{m^4_Z}+64\frac{m^6_l}{m^6_Z} \right)m^6_ZE_0(1)\, ,
\end{eqnarray}

\begin{eqnarray}
	{\cal A}^-_{ZZZ}=\frac{1}{6}\left[3B_0(1)+2B_0(2)-5B_0(3)\right]+\frac{3}{2}\left(1+\frac{1}{3}\frac{m^2_l}{m^2_Z}\right)m^2_ZC_0(1)\nonumber\\
	\qquad\quad-\frac{5}{3}\left( 1+\frac{7}{10}\frac{m^2_l}{m^2_Z}\right)m^2_ZC_0(2)
	+\frac{1}{3}m^2_ZC_0(3)\nonumber\\
	\qquad\quad+\frac{3}{2}\left(1+\frac{2}{3}\frac{m^2_l}{m^2_Z}-\frac{4}{3}\frac{m^4_l}{m^4_Z}\right)m^4_ZD_0(1)-\frac{5}{6}\left(1+\frac{7}{5}\frac{m^2_l}{m^2_Z} \right)m^4_ZD_0(2)\nonumber \\
	\qquad\quad+\frac{1}{2}\left(1+\frac{m^2_l}{m^2_Z}-4\frac{m^4_l}{m^4_Z}\right)m^6_ZE_0(1)\, .
\end{eqnarray}

\begin{eqnarray}
\fl	{\cal A}_{\gamma \gamma Z}=\left(\frac{m_Z}{m_l}\right)^2\Bigg\{ -\frac{3}{2}\left(1+3\frac{m^2_l}{m^2_Z}\right)+\frac{1}{4}\left[11B_0(3)-6B_0(2)-5B_0(1)\right]\nonumber \\
\fl	\qquad\qquad+\frac{m^2_l}{m^2_Z}\left[13B_0(3)-6B_0(2)-7B_0(1)\right]-\left(\frac{5}{4}+6\frac{m^2_l}{m^2_Z}-12\frac{m^4_l}{m^4_Z}\right) m^2_Z C_0(1)\Bigg\}\, .
\end{eqnarray}

\begin{eqnarray}
\fl	{\cal A}_{\gamma ZZ}=\left(\frac{m_Z}{m_l}\right)^2\Bigg\{ -\frac{3}{2}\left(1-\frac{m^2_l}{m^2_Z}\right) +\frac{1}{2}\left[11B_0(3)-6B_0(2)-5B_0(1) \right]
	\nonumber\\
\fl	\qquad\qquad+\frac{m^2_l}{m^2_Z}\left[13B_0(3)-6B_0(2)-7B_0(1)\right]-\frac{3}{4}\left(5+16\frac{m^2_l}{m^2_Z}-16\frac{m^4_l}{m^4_Z}\right)m^2_ZC_0(1)\nonumber\\
\fl	\qquad\qquad+\frac{1}{4}\left(11+52\frac{m^2_l}{m^2_Z}\right)m^2_ZC_0(2)-\left(\frac{5}{4}+6\frac{m^2_l}{m^2_Z}-12\frac{m^4_l}{m^4_Z}\right)m^4_ZD_0(1)\Bigg\}\, .
\end{eqnarray}

\begin{eqnarray}
\fl	{\cal A}_{\gamma Z \gamma}=\left(\frac{m_Z}{m_l}\right)^2\Bigg[-\frac{3}{2}\left(1+3\frac{m^2_l}{m^2_Z}\right)+\frac{1}{4}\left[11B_0(3)-6B_0(2)-5B_0(1) \right]
	\nonumber\\
\fl	\qquad\qquad+\frac{m^2_l}{m^2_Z}\left[13B_0(3)-6B_0(2)-7B_0(1) \right]-\left(\frac{5}{4}+6\frac{m^2_l}{m^2_Z}+12\frac{m^4_l}{m^4_Z}\right)m^2_ZC_0(1)\Bigg]\, .
\end{eqnarray}

\begin{equation}
	{\cal A}_{Z\gamma Z}=\left((g^l_A)^2+(g^l_V)^2\right){\cal A}^+_{Z\gamma Z}+\left((g^l_A)^2-(g^l_V)^2\right){\cal A}^-_{Z\gamma Z}\, ,
\end{equation}
with
\begin{eqnarray}
\fl {\cal A}^+_{Z\gamma Z}=\left(\frac{m_Z}{m_l}\right)^2\Bigg[-\frac{3}{2}\left(1-\frac{m^2_l}{m^2_Z}\right)+\frac{1}{2}\left[B_0(3)-6B_0(2)+5B_0(1)  \right]\nonumber\\
\fl \quad\quad+\frac{1}{3}\frac{m^2_l}{m^2_Z}\left[19B_0(3)-10B_0(2)-9B_0(1)\right]-\frac{1}{4}\left(15+16\frac{m^2_l}{m^2_Z}-64\frac{m^4_l}{m^4_Z}\right)m^2_ZC_0(1)\nonumber\\
\fl \quad\quad+\frac{1}{12}\left(33+76\frac{m^2_l}{m^2_Z}
	-112\frac{m^4_l}{m^4_Z}\right)m^2_ZC_0(2)-\frac{1}{4}\left(5+8\frac{m^2_l}{m^2_Z}-64\frac{m^4_l}{m^4_Z}+64\frac{m^6_l}{m^6_Z}\right)D_0(1)\Bigg],\nonumber\\
\end{eqnarray}

\begin{eqnarray}
	{\cal A}^-_{Z\gamma Z}=\frac{1}{3}\left[12B_0(1)+8B_0(2)-20B_0(3)\right]+4\left(2+\frac{m^2_l}{m^2_Z}\right)m^2_ZC_0(1)\nonumber \\
	\qquad\quad-\frac{5}{3}\left(1+\frac{7}{5}\frac{m^2_l}{m^2_Z}\right)m^2_ZC_0(2)+4\left(1+\frac{m^2_l}{m^2_Z}-4\frac{m^4_l}{m^4_Z}\right)m^4_ZD_0(1)\, .
\end{eqnarray}

\begin{eqnarray}
	{\cal A}_{(1+2)}&=&\frac{5}{3}\left[B_0(7)-B_0(8)\right]-2\left(1-\frac{5}{3}\frac{m^2_l}{m^2_W}\right)m^2_WC_0(8)\nonumber \\
	&&+\frac{11}{3}\left(1-\frac{m^2_l}{m^2_W}\right)m^2_WC_0(9)+2\left(1-\frac{m^2_l}{m^2_W}\right)^2m^4_WD_0(5)\, .
\end{eqnarray}

\begin{eqnarray}
	{\cal A}_{(3+4+5)}=-\frac{3}{8}+\frac{1}{64}\left(99+121\frac{m^2_l}{m^2_W}\right)\left[B_0(8)-B_0(7)\right]\nonumber\\
	\qquad\qquad-\frac{1}{12}\left(27+64\frac{m^2_l}{m^2_W}+
	37\frac{m^4_l}{m^4_W}\right)m^2_WC_0(8)\nonumber \\
	\qquad\qquad+\frac{1}{48}\left(189+362\frac{m^2_l}{m^2_W}+121\frac{m^4_l}{m^4_W}\right)m^2_WC_0(9)\nonumber \\
	\qquad\qquad+\frac{1}{48}\left(153+349\frac{m^2_l}{m^2_W}+235\frac{m^4_l}{m^4_W}-161\frac{m^6_l}{m^6_W}\right)m^4_WD_0(5)\nonumber \\
	\qquad\qquad-\frac{1}{16}\left(15+41\frac{m^2_l}{m^2_W}+57\frac{m^4_l}{m^4_W}-33\frac{m^6_l}{m^6_W}\right)m^4_WD_0(6)\nonumber \\
	\qquad\qquad+\frac{1}{16}\left(5+12\frac{m^2_l}{m^2_W}-14\frac{m^4_l}{m^4_W}-20\frac{m^6_l}{m^6_W}-11\frac{m^8_l}{m^8_W}\right)m^6_WE_0(3)\, .
\end{eqnarray}

In the above expressions, $B_0(i)$, $C_0(i)$, $D_0(i)$, and $E_0(i)$ are PV scalar functions of two, three, four, and five points, respectively, which are listed along with their solutions in Appendix C. All these PV functions have relatively simple solutions, so the various ${\cal A}_i$ functions can be expressed in terms of elementary functions. We will write the analytical solutions of these form factors together with the first terms of their corresponding power series around $x = 0$:

\begin{equation}
	{\cal A}_{\gamma \gamma \gamma}=\frac{21}{2}+4 \log\left(\frac{m^2_\gamma}{m^2_l}\right)\, .
\end{equation}
The ${\cal A}_{ZZZ}$ amplitude can be written as follows:
\begin{equation}
	{\cal A}_{ZZZ}=\left(\left(g^l_A\right)^2+ \left(g^l_V\right)^2\right)f^+_{ZZZ}(x)+\left(\left(g^l_A\right)^2- \left(g^l_V\right)^2\right)f^-_{ZZZ}(x)\, ,
\end{equation}
where $x=\frac{m_l}{m_Z}$. In addition,
\begin{eqnarray}	\fl f^+_{ZZZ}(x)=\frac{1}{(1-4x^2)^2}\left(\frac{3}{x^2}-\frac{463}{48}-\frac{347}{6}x^2+\frac{551}{3}x^4-8 x^6\right)+\left(\frac{3}{x^4}+\frac{143}{24x^2}-\frac{49}{6}\right)\log(x)\nonumber \\
\fl \qquad\qquad\,\,+\frac{1}{(1-4x^2)^{\frac{5}{2}}}\left(\frac{3}{x^4}-\frac{577}{24x}+\frac{37}{2}+230x^2
	-\frac{1330}{3}x^4+40x^6\right)\nonumber\\
\fl\qquad\qquad \quad\times \log\left(\frac{\sqrt{1+2x}+\sqrt{1-2x}}{\sqrt{1+2x}-\sqrt{1-2x}}\right)\, ,
\end{eqnarray}

\begin{equation}
f^+_{ZZZ}(x)=\frac{187}{48}+\frac{180}{48}\log(x)+O(x^2)\, ,
\end{equation}

\begin{eqnarray}
\fl f^-_{ZZZ}(x)=\frac{1}{(1-4x^2)^2}\left[-4+25x^2-22x^4-8x^6-\left(1+\frac{4}{x^2}\right)\log(x)\right]\nonumber \\
\fl \qquad\qquad+\frac{1}{(1-4x^2)^{\frac{5}{2}}}\left(-\frac{4}{x^2}+7-110x^2+50x^4+40x^6 \right)\log\left(\frac{\sqrt{1+2x}+\sqrt{1-2x}}{\sqrt{1+2x}-\sqrt{1-2x}}\right)\, ,\nonumber\\
\end{eqnarray}

\begin{equation}
	f^-_{ZZZ}=x^2\left(32+120\log(x)\right)   \, .
\end{equation}
Note that ${\cal A}_{ZZZ}$ tends to infinity logarithmically in the limit $x\to 0$. This behavior tells us that the result is much more sensitive to light lepton masses. It is worth mentioning that there are subtle cancellations in these functions.\\

The ${\cal A}_{\gamma \gamma Z}$ amplitude can be written as follows:
\begin{eqnarray}
	{\cal A}_{\gamma \gamma Z}=\frac{43}{2}+\frac{4}{x^2}-4\left(6-\frac{1}{x^4}-\frac{4}{x^2}\right)\log\left(x
	\right)\nonumber\\
	\qquad\quad+\frac{4}{\sqrt{1-4x^2}}\left(\frac{1}{x^4}+\frac{2}{x^2}-16\right)\log\left(\frac{\sqrt{1+2x}+\sqrt{1-2x}}{\sqrt{1+2x}-\sqrt{1-2x}}\right)\, ,
\end{eqnarray}

\begin{equation}
 {\cal A}_{\gamma \gamma Z}=-\frac{1}{2} -\frac{40}{3}x^2 +O(x^4)\, .
\end{equation}

In the case of the ${\cal A}_{\gamma Z Z}$ amplitude, we have
\begin{eqnarray}
\fl	{\cal A}_{\gamma ZZ}=\frac{1}{1-4x^2}\left(\frac{12}{x^2}+\frac{3}{2}-134x^2\right)-4\left(6-\frac{3}{x^4}-\frac{8}{x^2}\right)\log(x)\nonumber \\
\fl \qquad\quad+\frac{4}{(1-4x^2)^{\frac{3}{2}}}\left(\frac{3}{x^4}-\frac{10}{x^2}-36+96x^2\right)\log\left(\frac{\sqrt{1+2x}+\sqrt{1-2x}}{\sqrt{1+2x}-\sqrt{1-2x}}\right)\, ,
\end{eqnarray}

\begin{equation}
 {\cal A}_{\gamma Z Z}=-\frac{1}{2}+x^4\left(\frac{241}{3}+56\log(x)\right)+O(x^6)\, .
\end{equation}

The ${\cal A}_{\gamma Z \gamma}$ amplitude can be expressed as follows:

\begin{eqnarray}
	{\cal A}_{\gamma Z \gamma}=\frac{43}{2}+\frac{4}{x^2}+4\left(\frac{1}{x^4}+\frac{4}{x^2}\right)\log(x)\nonumber\\
	\qquad\quad+\frac{4}{\sqrt{1-4x^2}}\left(\frac{1}{x^4}+
	\frac{2}{x^2}-10\right)\log\left(\frac{\sqrt{1+2x}+\sqrt{1-2x}}{\sqrt{1+2x}-\sqrt{1-2x}}\right)\, ,
\end{eqnarray}

\begin{equation}
 {\cal A}_{\gamma Z \gamma}=-\frac{1}{2}-x^2\left(\frac{112}{3}+48\log(x)\right)+O(x^4)\, .
\end{equation}

The ${\cal A}^\pm_{Z\gamma Z}$ amplitudes are given by:
\begin{eqnarray}
\fl 	{\cal A}^+_{Z\gamma Z
	}=\frac
	{169}{6}+\frac{12}{x^2}+24x^2+\frac{128x^4}{1-4x^2}+4\left(\frac{3}{x^4}+\frac{8}{3x^2}-8\right)\log(x)\nonumber \\
\fl\qquad\quad+\frac{4}{3\left(1-4x^2\right)^{\frac{3}{2}}}\left(\frac{9}{x^4}-\frac{46}{x^2}-18+228x^2-80x^4\right)\log\left(\frac{\sqrt{1+2x}+\sqrt{1-2x}}{\sqrt{1+2x}-\sqrt{1-2x}}\right)\, , \nonumber\\
\end{eqnarray}

\begin{equation}
 {\cal A}^+_{ Z \gamma Z}=-\frac{1}{2}+x^4\left(\frac{319}{9}+\frac{88}{3}\log(x)\right)+O(x^6)\, .
\end{equation}
On the other hand,
\begin{eqnarray}
\fl {\cal A}^-
	_{Z\gamma Z}=\frac{8x^2}{3\left(1-4x^2\right)}\left(-8+17x^2+12x^4\right)-\left(\frac{49}{3}+x^2\right)\log(x)\nonumber \\
\fl\qquad\quad-\frac{1}{3\left(1-4x^2\right)^{\frac{3}{2}}}\left(49-291x^2+246x^4+152x^6\right)\log\left(\frac{\sqrt{1+2x}+\sqrt{1-2x}}{\sqrt{1+2x}-\sqrt{1-2x}}\right)\, ,
\end{eqnarray}

\begin{equation}
	{\cal A}^-_{ Z \gamma Z}=-5x^2+O(x^4)\, .
\end{equation}

Finally, the total $W$ gauge boson contribution ${\cal A}_{WWW}={\cal A}_{(1+2)}+{\cal A}_{(3+4+5)}$ can be written as follows:
\begin{eqnarray}
	{\cal A}_{WWW}&=&\frac{1}{(1-y^2)^2}\left(\frac{223}{32}-\frac{567}{64}y^2+\frac{417}{32}y^4-\frac{119}{8}y^6+\frac{29}{8}y^8+\frac{55}{192}y^{10}\right)\nonumber \\
	&&+\left(\frac{1483}{192 y^2}+\frac{179}{96}+\frac{847}{192}y^2\right)\log\left(1-y^2\right)\, ,
\end{eqnarray}
where $y=\frac{m_l}{m_W}$. In the $y\to 0$ limit, we have

\begin{equation}
	{\cal A}_{WWW}(y)=-\frac{145}{192}+\frac{611}{192}y^2+O(y^4)\, .
\end{equation}

It is important to note that all form factors ${\cal A}_i$ tend to a constant value other than zero in the limit $x,y \to 0$, with the exception of ${\cal A}_{ZZZ } $, which tends to infinity logarithmically in that limit. This strong non-decoupling effect dominates the contribution to the MDM.

\section*{Appendix C: Passarino-Veltman Scalar Functions}
\label{AC}
\label{PVSF}In this appendix, we present a list of the diverse Passarino-Veltman scalar functions that appear in the calculations. Using the notation of the FeynCalc program~\cite{FCP}, these functions are:
\begin{eqnarray}
	A_0(1)&=&A_0(m^2_l)\, , \\
	A_0(2)&=&A_0(m_Z^2)\, , \\
	B_0(1)&=&B_0(0,m_l^2,m_l^2)\, , \\
	B_0(1,q^2)&=&B_0(q^2,m_l^2,m_l^2)\, , \\
	B_0(2)&=&B_0(0,m_Z^2,m_Z^2)\, ,\\
	B_0(3)&=&B_0(m_l^2,m^2_Z,m_l^2)\, ,\\
	C_0(1)&=&C_0(m^2_l,m^2_l,0,m^2_l,m^2_Z,m^2_l)\, ,\\
	C_0(1,q^2)&=&C_0(m^2_l,m^2_l,q^2,m^2_l,m^2_Z,m^2_l)\, ,\\
	C_0(2)&=&C_0(m^2_l,m^2_l,0,m^2_Z,m^2_l,m^2_Z)\, , \\
	C_0(3)&=&C_0(0,0,0,m^2_Z,m^2_Z,m^2_Z)\, ,\\
	D_0(1)&=&D_0(m^2_l,m^2_l,m^2_l,m^2_l,0,0,m^2_Z,m^2_l,m^2_Z,m^2_l)\, ,\\
	D_0(1,q^2)&=&D_0(m^2_l,m^2_l,m^2_l,m^2_l,0,q^2,m^2_Z,m^2_l,m^2_Z,m^2_l)\, ,\\
	D_0(2)&=&D_0(m^2_l,m^2_l,0,0,0,m^2_l,m^2_Z,m^2_l,m^2_Z,m^2_Z)\, ,
\end{eqnarray}
\begin{equation}
	E_0=E_0(0,0,m^2_l,0,m^2_l,0,m^2_l,m^2_l,m^2_l,m^2_l,m^2_Z,m^2_Z,m^2_Z,m^2_l,m^2_l)\, ,
\end{equation}

\begin{eqnarray}
	B^W_0(1)&=&B_0(0,m^2_W,m^2_W)\, , \\
	B^W_0(1,q^2)&=&B_0(q^2,m^2_W,m^2_W)\, , \\
	B^W_0(2)&=&B_0(m^2_l,0,m^2_W)\, , \\
	C^W_0(3)&=&C_0(0,0,0,m^2_W,m^2_W,m^2_W)\, ,\\
	C^W_0(3,q^2)&=&C_0(0,q^2,q^2,m^2_W,m^2_W,m^2_W)\, ,\\
	C^W_0(4)&=&C_0(0,m_l^2,m_l^2,m^2_W,m^2_W,0)\, ,\\
	C^W_0(4,q^2)&=&C_0(q^2,m_l^2,m_l^2,m^2_W,m^2_W,0)\, ,\\
	D^W_0(1)&=&D_0(0,m^2_l,m^2_l,0,m^2_l,0,m^2_W,m^2_W,0,m^2_W)\, ,\\
	D^W_0(1,q^2)&=&D_0(0,m^2_l,m^2_l,q^2,m^2_l,q^2,m^2_W,m^2_W,0,m^2_W)\, .
\end{eqnarray}
All these functions can be expressed in terms of elementary functions. The solutions of the $A_0(m^2)$ function is given by:
\begin{equation}
	A_0(m^2)=m^2\left(\Delta +1-\log\left(\frac{m^2}{\mu^2}\right)\right)\, ,
\end{equation}
where $\Delta=\frac{2}{\epsilon}-\gamma_E+\log\left(4\pi\right)$, with $\gamma_E$ the Euler-Mascheroni constant and $\mu$ the scale of the dimensional regularization scheme. On the other hand, functions of the type $B_0(1)$ or $B_0(2)$ have solutions of the way:
\begin{equation}
	B_0(0,m^2,m^2)=\Delta-\log\left(\frac{m^2}{\mu^2}\right)\, .
\end{equation}
In addition,
\begin{equation}
	B^W_0(2)=\Delta+2-\log\left(\frac{m^2_W}{\mu^2}\right)+\frac{m^2_l-m^2_W}{m^2_l}\log\left(\frac{m^2_W}{m^2_W-m^2_l}\right)\, .
\end{equation}

The $B_0(3)$ function has a slightly more complicated solution,
\begin{eqnarray}
	B_0(3)&=&\Delta +2-\frac{1}{2}\log\left(\frac{m^2_Z}{\mu^2}\right)-\frac{1}{2}\log\left(\frac{m^2_l}{\mu^2}\right)+
	\frac{1}{2}\left(1-\frac{m^2_Z}{m^2_l}\right)\log\left(\frac{m^2_Z}{m^2_l}\right)\nonumber \\
	&&+\frac{m^2_Z}{m^2_l}\sqrt{1-\frac{4m^2_l}{m^2_Z}}\log\left(\frac{\sqrt{1+\frac{2m_l}{m_Z}}+
		\sqrt{1-\frac{2m_l}{m_Z}}}{\sqrt{1+\frac{2m_l}{m_Z}}-\sqrt{1-\frac{2m_l}{m_Z}}}\right)\, .
\end{eqnarray}
As far as the solutions of the $C_0$ functions are concerned, they are given by:
\begin{eqnarray}
	C_0(1)&=&\frac{2+2B_0(2)-B_0(1)-B_0(3)}{m^2_Z-4m^2_l}\, ,\\
	C_0(2)&=&\frac{\partial}{\partial(m^2_Z)}B_0(3)\, ,\\
	C_0(3)&=&-\frac{1}{2m^2_Z}\, .
\end{eqnarray}
On the other hand, the $D_0$ functions have the following solutions:
\begin{eqnarray}
	D_0(1)&=&\frac{\partial}{\partial(m^2_Z)} C_0(1)\, , \\
	D^W_0(1)&=&\frac{1}{2 m^2_W(m^2_W-m^2_l)}\, ,\\
	D_0(2)&=&\frac{1}{2}\frac{\partial^2}{\partial(m^2_Z)^2}B_0(3)\, .
\end{eqnarray}
Finally, the solution of the $E_0$ function is given by:
\begin{equation}
	E_0=\frac{1}{2}\frac{\partial^2}{\partial(m^2_Z)^2}C_0(1)\, .
\end{equation}

\ack
We acknowledge financial support from CONAHCYT (Consejo Nacional de Humanidades, Ciencias y Tecnolog\'\i as) and SNII (Sistema Nacional de Investigadoras e Investigadores) (M\' exico).

\end{document}